\documentclass[fleqn,usenatbib]{mnras}

\usepackage{algorithm,algorithmic}
\usepackage{url}

\usepackage{natbib}
\usepackage[T1]{fontenc}
\usepackage{graphicx}	% Including figure files
\usepackage{amsmath}	% Advanced maths commands
\usepackage{amssymb}	% Extra maths symbols

\usepackage{graphicx}% Include figure files
\usepackage{subcaption}
\usepackage{dcolumn}% Align table columns on decimal point
\usepackage{bm}% bold math
\usepackage{hyperref}% add hypertext capabilities
\usepackage{cleveref}
\usepackage[dvipsnames]{xcolor}

\newcommand{\hmpc}{{$h^{-1}$Mpc}}
\definecolor{ForestGreen}{rgb}{0.2,0.6,0.2}

\newcommand{\muscle}{{\scshape muscle}}

\crefformat{figure}{Fig.~#2#1#3}
\crefformat{equation}{Eq.~(#2#1#3)}

\title[Simulations with translation process theory]{Beyond the Lognormal Approximation: a General Simulation Scheme}

\author[F. Tosone et al.]{Federico Tosone,$^{1,2}$\thanks{E-mail: federico.tosone@roma2.infn.it}
Mark C.\ Neyrinck,$^{3,4,5}$
Benjamin R. Granett,$^{6,7}$
\newauthor
Luigi Guzzo,$^{6,7,8}$
Nicola Vittorio$^{1,2}$
\\
$^{1}$Dipartimento di Fisica, Universit\`{a} di Roma Tor Vergata, Via della Ricerca Scientifica 1, I-00133, Roma, Italy\\
$^{2}$INFN Sezione di Roma Tor Vergata, Via della Ricerca Scientifica 1, I-00133, Roma, Italy\\
$^{3}$Ikerbasque, the Basque Foundation for Science\\
$^{4}$Dept.\ of Theoretical Physics, University of the Basque Country, Bilbao, Spain\\
$^{5}$Donostia International Physics Center, San Sebasti\'{a}n, Spain\\
$^{6}$Universit\`{a} degli Studi di Milano, via G. Celoria 16, 20133 Milano, Italy\\
$^{7}$INAF - Osservatorio Astronomico di Brera, via Brera 28, 20122 Milano, Italy\\
$^{8}$INFN Sezione di Milano, Via G. Celoria 16, 20133, Milano, Italy
}

\date{Accepted XXX. Received YYY; in original form ZZZ}

\pubyear{2015}

\begin{document}
\label{firstpage}
\pagerange{\pageref{firstpage}--\pageref{lastpage}}
\maketitle

\begin{abstract}
	We present a public code to generate random fields with an arbitrary probability distribution function (PDF) and an arbitrary correlation function. The algorithm is cosmology-independent, applicable to any stationary stochastic process over a three dimensional grid. We implement it in the case of the matter density field, showing its benefits over the lognormal approximation, which is often used in cosmology for generation of mock catalogues. We find that the covariance of the power spectrum from the new fast realizations is more accurate than that from a lognormal model. As a proof of concept, we also apply the new simulation scheme to the divergence of the Lagrangian displacement field. We find that information from the correlation function and the PDF of the displacement-divergence provides modest improvement over other standard analytical techniques to describe the particle field in the simulation. This suggests that further progress in this direction should come from multi-scale or non-local properties of the initial matter distribution.
\end{abstract}


%%%%%%%%%%%%%%%%%%%%%%%%%%%%%%%%%%%%%%%%%%%%%%%%%%

%%%%%%%%%%%%%%%%% BODY OF PAPER %%%%%%%%%%%%%%%%%%

%\begin{abstract}
%	%\noindent
%\end{abstract}

\maketitle

%\tableofcontents

\section{Introduction}

%\noindent
Analyses of the large-scale structure of the Universe have made huge progress during the past two decades. An unprecedented wealth of data was provided by redshift surveys of larger and larger volumes out to $z\sim 1$ -- as, e.g., 2dFGRS \citep{Colless03}, SDSS/BOSS \citep{Alam17} WiggleZ \citep{Blake11b} and VIPERS \citep{Guzzo14} -- with a further order-of-magnitude increase expected soon from new projects like DESI \citep{DESI} and Euclid \citep{Laureijs11}. In this scenario, the availability of large numbers of simulated mock redshift surveys or weak-lensing maps has become a crucial requirement, as to assess both precision and accuracy of the recovered cosmological parameters. This is producing a huge effort in the community, with the goal of improving fast simulation schemes as an alternative to full $N$-body. This is motivated not only to save computational costs, but also because an improvement of these techniques could help to shed light on the statistical properties of the Cold Dark Matter (CDM) density field.

One widespread technique for the production of such catalogues is based on the observation that the evolution of the density field can be approximated with an exponential growth starting from Gaussian initial conditions, thus resulting in a lognormal distribution \citep{Coles91}. Lognormal fields are very convenient as they can be simulated having an arbitrary correlation function that can be always written in terms of the correlation of an underlying Gaussian field, and a PDF resembling that of the density field (e.g. \cite{Agrawal17,Favole20}). Nonetheless, there is growing evidence that the PDF of the density field does not exactly follow a lognormal distribution, and could have a significant effect on the estimate of cosmological parameters \citep{Uhlemann19}. With the exception of the lognormal case, the simulation of arbitrary random fields has made little progress in cosmology. The main purpose of this paper is to introduce an improvement over the limitation of the lognormal approximation, with a scheme that allows for the simulation of a random field with an arbitrary PDF and arbitrary two-point correlation structure. Similarly to the lognormal case, where a Gaussian field is mapped onto a lognormal one, the new algorithm optimizes for the Gaussian power spectrum to be used to generate many realizations of a field that can be mapped onto a target field. The underlying idea is based on the so-called translation process theory, detailed in \cref{sec:TP}. The algorithm we implement was proposed by \cite{Shields11} in the context of one-dimensional time-series simulations, and it is dubbed Iterative Translation Approximation Method (ITAM). We extend and adapt this scheme to the case of interest of 3D simulation grids. It can be regarded as a generalization of the lognormal case to distributions where no analytical formulae are available. Unlike previous schemes implemented in cosmology, the algorithm is not based on the realizations of the field itself on the grid, but it works at the level of the correlation function, thus saving computational cost and being fairly independent of the resolution, which can be chosen appropriately. Our results are reported in \cref{sec:Results}, where we also compare the covariance matrix obtained from these improved realizations against those of the lognormal model and from $N$-body simulations. The method preserves the overall structure of the covariance matrix, as the lognormal field does, with a marginal improvement over it. We also find an analytical justification of why this structure is preserved, by arguing that it is set by the an exponential growth of structures.

In the last part of this work, \cref{sec:lagscheme}, as a further test of the algorithm we apply it to another random field, the Lagrangian displacement-divergence \citep{Bouchet1995}. This is the fundamental field for simulation schemes based on Lagrangian techniques \citep{Monaco16}, which are often used for the fast generation of mock catalogues \citep{Blot19}. Indeed they are theoretically sound, since they are based on approximate solutions for the CDM evolution, but they have some limitations in reproducing the property of the density field, such as the the amount of correlation at small scales \citep{Chuang15}, the skewness of the PDF of the matter distribution \citep{Neyrinck15}, and the cross correlation with the actual dark matter evolution that is far from ideal \citep{Munari17}. We use ITAM to simulate a displacement-divergence field with a given power spectrum and PDF, accurately reproducing the $N$-body results. We assess how much information the PDF and the power spectrum convey and compare it with other analytical techniques used in the literature. We find that the displacement field generated by ITAM performs similarly to other standard techniques, with a minor improvement manifested in more-collapsed filaments, and walls. These results suggest that more accurate Lagrangian schemes are accessible if one resorts to non-local transformations of the initial density field or higher order polyspectra.

\section{Translation Process Theory}\label{sec:TP}
\subsection{Basics}
It is now well-known that the initial conditions of the Universe closely resemble a Gaussian field. This distribution, under the influence of gravity, gives rise to complex features that are not completely characterized by the two-point correlation function alone \citep{Bernardeau01}. The evolved matter density field is nonlinear, referring both to the fact that it cannot be computed from linear perturbation theory, and that it cannot be obtained as the linear superposition of Gaussian processes. An important insight in the statistics of matter clustering was provided by the observation that the distribution of galaxies follows approximately a lognormal distribution. A lognormal field can be extrapolated as a solution from the continuity equation of the CDM evolution under the Zel'dovich approximation (ZA) \citep{Coles91}. Several works showed that applying a logarithmic transform to the nonlinear density field renders it more Gaussian at the level of the one-point distribution function \citep{Colombi94} and it can improve the constraints on cosmological parameters \citep{Neyrinck09,Repp17}. A similar effect is obtained by Gaussianizing the field, namely by rank ordering the nonlinear density at the voxel level, and then mapping it onto a Gaussian PDF \citep{Weinberg92,Neyrinck11}, without making the assumption of lognormality for the density field.

Based on these evidences, both of these approaches have been extensively used in the literature in the inverse sense: an initial Gaussian field is transformed into a lognormal field \citep{Agrawal17} or to arbitrary distributions \citep{Shirasaki17}. Fast simulations generated with such a heuristc approach have limitations. In the former case they have the exact correlation function but incorrect PDF, and in the latter they have the exact PDF but incorrect correlation function. To be more formal, let us consider a Gaussian stationary stochastic process $\delta_g(\pmb{x},\tau)$. An example of such a process is the matter density field, where stationarity is guaranteed by translational invariance. $\delta_g(\pmb{x},\tau)$ can be mapped onto the target non-Gaussian variable $\delta_{ng}(\pmb{x},\tau)$ through a monotonic transformation
\begin{equation}
	\delta_{ng}(\pmb{x},\tau) = g[\delta_g(\pmb{x},\tau)].
\end{equation}

This kind of transformation is \textit{local}, as it is a one-point mapping from a distribution to another that depends only on the local value of the field, and does not explicitly depend on its coordinates. The transformation of a Gaussian variable by means of a local nonlinear transformation has been dubbed elsewhere as a translation process \citep{Grigoriu84}, and it has the property that the new correlation structure is also translation-invariant and thus it is still a stationary stochastic process.
The most precise one-point mapping that performs the transformation from $\delta_g$ to $\delta_{ng}$ is obtained by matching their cumulative distribution functions (CDFs) $\mathcal{F}_g$ and $\mathcal{F}_{ng}$, namely:
\begin{equation}\label{eq:g}
\delta_{ng} = g[\delta_g] \equiv \mathcal{F}^{-1}_{ng} \left[ \mathcal{F}_g [\delta_g] \right],
\end{equation}
where $\mathcal{F}^{-1}$ denotes the inverse CDF. It is standard practice to use this relation to transform between two distributions. It simply consists of matching the rank-ordering of $\delta_{ng}$ to the one of $\delta_g$, and it can be generalized to arbitrary distributions (see e.g. \cite{Leclercq13}). However, there is no control over the resulting correlation structure of the nonlinear field, as we are going to address now. We start by computing the expected first two moments resulting from a monotonic transformation of this kind \citep{Grigoriu95}:
\begin{equation}
	\mu = \int_{-\infty}^{+\infty} g(\delta_g) \phi(\delta_g) d\delta_g,
\end{equation}
\begin{equation}
	\sigma^2 = \int_{-\infty}^{+\infty} (g(\delta_g)-\mu)^2 \phi(\delta_g) d\delta_g,
\end{equation}
and the correlation function
\begin{equation}\label{eq:Grigoriu}
	\begin{split}
		\xi(r) &\equiv \langle \delta_{ng}(\pmb{x})\delta_{ng}(\pmb{x}+\pmb{r}) \rangle =\\ 
		&\int_{-\infty}^{+\infty} \int_{-\infty}^{+\infty} (g(\delta_g)-\mu) (g(\delta'_g)-\mu) \phi(\delta_g,\delta'_g,\rho(r)) d\delta_g d\delta'_g,
	\end{split}
\end{equation}
where $\delta_{g}=\delta_{g}(\pmb{x})$, $\delta'_{g}=\delta'_{g}(\pmb{x}+\pmb{r})$ and we made use of the univariate and bivariate normal distributions
\begin{equation}
	\phi(x)= (2 \pi)^{-1/2} \exp{(-x^2/2)},
\end{equation}
\begin{equation}
	\phi(x,y,\rho(r))= \frac{1}{\sqrt{2 \pi(1-\rho^2(r))}} \exp{ \left(\frac{-(x^2+y^2-2 \rho^2(r) x y)}{2 \pi(1-\rho^2(r)) } \right)}.
\end{equation}

\Cref{eq:Grigoriu} explicitly connects the correlation function $\rho(r)=\xi_g(r)/\xi_g(0)$ of the pre-translation Gaussian field $\delta_g$ with the new correlation function $\xi_{ng}(r)$ of $\delta_{ng}$. This equation can always be used in a forward sense, i.e. for a given monotonic transform $g$ and a Gaussian variable $\delta_g$ we can compute the new correlation function. In practice, we would like to solve the \textit{inverse problem}: we want to find the linear correlation $\xi_g$ which allows for the generation of the pre-translation field $\delta_g$ that, under transform \cref{eq:g}, is mapped onto a field $\delta_{ng}$ with prescribed correlation $\xi_{ng}$. While the new field $\delta_{ng}$ has by construction the exact PDF, there is no reason why the resulting correlation matches the desired target. This is possible only if we can solve \cref{eq:Grigoriu} for the correlation structure $\xi_g$ of the pre-translation field in terms of $\xi_{ng}$. To clarify, one explicit example where this is instead possible is the case of the lognormal density field. The monotonic transformation in this case is written explicitly as
\begin{equation}\label{eq:ln}
	\delta_{L}(\pmb{x}) + 1 = e^{\delta_{g}(\pmb{x})-\sigma^2/2},
\end{equation}
where $\sigma$ is the variance of $\delta_{g}$, which guarantees that the mean of $\delta_{L}$ is zero. With this transformation, \cref{eq:Grigoriu} can be solved exactly, yielding
\begin{equation}\label{eq:ln2}
	\xi^{L}_{ng}(r) = e^{\xi_g(r)}-1.
\end{equation}

The advantage of a lognormal field is clear: from the previous equation we can always solve for $\xi_{g}$ for any given $\xi^{L}_{ng}$ (if $1+\xi^{L}_{ng}>0$). 

More recently, \cite{Bel16} (their appendix C) solved the inverse problem by Hermite-expanding the integral \cref{eq:Grigoriu} to linear order in $\xi_g$. This gives a linear relation $\xi_{ng} \propto \xi_{g}$, from which one can always find $\xi_{g}$. This approach is also the basis for the algorithm developed in \cite{Baratta20} for generating mock galaxy catalogues. While it is an excellent approximation to describe the non-Gaussian correlation smoothed at quasi nonlinear scales, it is not optimal if one is interested in the nonlinear regime (\cref{fig:correlations}). Deviations from the linear relation become apparent when considering a density field smoothed at scales smaller than $8$ \hmpc, when $\xi_{ng}(0)<1$. These deviations might be relevant at smoothing scales of the order of $5$ \hmpc.

Apart from these exceptions, one might be interested in the case where the translation transform is not known analytically, like in the case of the density field PDF. Even assuming that the one-point mapping is known and that \cref{eq:Grigoriu} can be solved analytically, it might still not possible to solve explicitly for the Gaussian correlation in a non-perturbative form. Therefore, the inverse problem can be seen as an optimization problem for the correlation structure $\xi_g$ of the pre-translation field. In the following we present an algorithm that can always perform this optimization to solve the inverse problem in a very general way.

\subsection{Iterative Translation Approximation Method}
In order to solve the inverse problem for the general case when no analytical solution is available, several numerical procedures were proposed in the past. The very first one was suggested by \cite{Yamazaki88}, which we comment briefly upon, since it is the only one that received some attention in cosmology. In particular the Yamazaki-Shinozuka (YS) algorithm has been discussed thoroughly in \cite{Vio01,Brown13}. The YS scheme is an iterative procedure that works directly on sample realizations of fields. Given a nonlinear sample realization of a field, one can measure the corresponding PDF, and consequently the translation transform $g$ of \cref{eq:g}. One then needs to initialize an arbitrary random Gaussian field, on which the transformation $g$ is applied. The translated field by construction has the prescribed one-point distribution, but it generally has a different power spectrum from the desired one, because of the arbitrariness of the pre-translation power spectrum.

In order to match a target power spectrum, one needs to optimize the pre-translation field so that it gets mapped more precisely onto the target nonlinear field. Optimizing the Gaussian field is a cumbersome task, and YS attempted to solve it by iteratively updating the amplitude of the modes of the pre-translation field by a factor proportional to the amplitude of the modes of the target field. The same phases of the zero-iteration Gaussian modes are used, and the newly generated Gaussian field gets translated and compared to the target field. This procedure is repeated until the amplitude of the modes of the translated field are close to the amplitude in the target field. Unfortunately, the YS optimization has an important limitation: the updating scheme based on the actual modes of the target realization introduces non-Gaussian correlations on the amplitudes of the pre-translation Gaussian modes. This causes the pre-translation field to be non-Gaussian \citep{Deodatis01}. Furthermore, working at the level of sample realizations of the fields makes the result realization-dependent. Many different refinements to ameliorate this problem have been explored, and they are reviewed and discussed in \cite{Bocchini08}. For the sake of this work, we instead focus on a recent scheme that elegantly addresses both problems.

This scheme was proposed in \cite{Shields11}, where the discussion is developed for time series simulations. We extend and adapt the suggested scheme to three-dimensional
grid simulations. The goal is the same as outlined in the YS scheme: we choose a target PDF $\mathcal{P}(\delta_{ng})$ and we initialize an arbitrary Gaussian power spectrum $P^{(i)}_g(k)$ for the zero iteration $(i=0)$. This spectrum can be converted into a correlation function by means of an Hankel transform:
\begin{equation}\label{eq:WK}
	\xi^{(i)}_g(r) = \int_0^{\infty} \frac{dk}{2 \pi^2} P^{(i)}_g(k) k^2 \frac{\sin{(kr)}}{kr}.
\end{equation}

This correlation function is normalized ($\rho_g=\xi_g(r)/\xi_g(0)$), and inserted into \cref{eq:Grigoriu} to compute the new correlation function $\xi_{ng}$. To perform the integration, we need the monotonic mapping \cref{eq:g}. This mapping is automatically set once $\mathcal{P}(\delta_{ng})$ has been specified, and in practice one does not need directly the PDF, but only the corresponding CDF. After computing $\xi_{ng}$, one applies the inverse transform:
\begin{equation}\label{eq:inverseWK}
	P^{(i)}_{ng}(k) = 4 \pi \int_0^{\infty} d r \xi^{(i)}_{ng}(r) r^2 \frac{\sin{(kr)}}{kr}.
\end{equation}

This power spectrum can be directly compared with the target power spectrum $P(k)$. If the power fails to converge, we update the Gaussian power spectrum to be used in the next iteration as
\begin{equation}\label{eq:update}
	P^{(i+1)}_g(k) = \left(\frac{P(k)}{P^{(i)}_{ng}(k)}\right)^{\beta} P^{(i)}_g(k).
\end{equation}

\begin{figure}
	\includegraphics[width=0.45\textwidth]{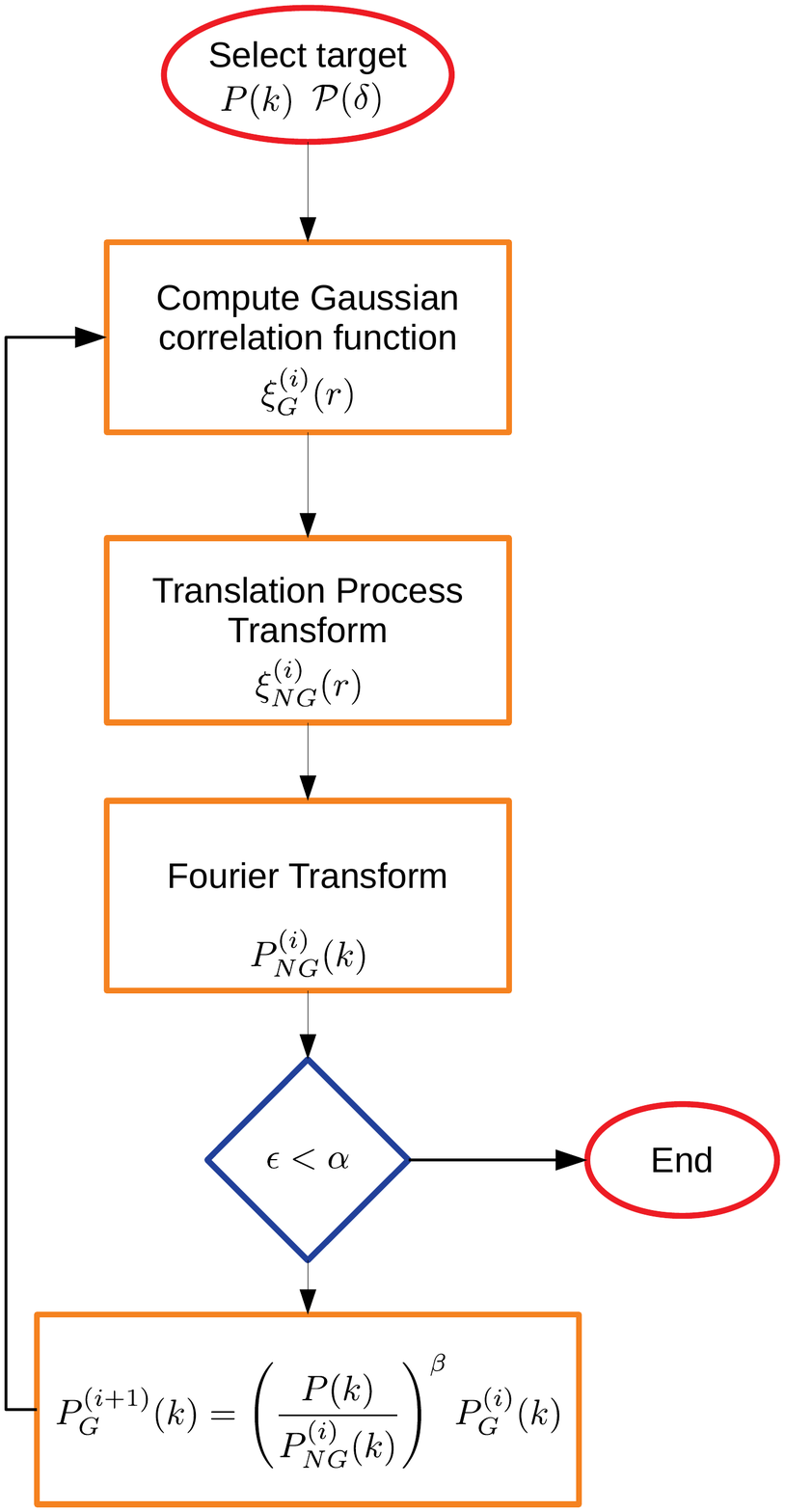}
	\caption{Diagrammatic representation of ITAM algorithm.\label{fig:FlowChart}}
\end{figure}

The scheme is summarized in \cref{fig:FlowChart}. The convergence criterion can be based on the relative difference between the target power spectrum $P(k)$ and the ITAM output $ P_{ng}(k)$ over all the modes of the box
\begin{equation}\label{eq:stop}
	\epsilon^{(i)} = \sqrt{ \frac{\sum_j \left(P^{(i)}_{ng}(k_j) - P(k_j) \right)^2 }{\sum_{j} P^2(k_{j}) } } < \alpha,
\end{equation}
where $\alpha$ and $\beta$ are arbitrary constants. This stopping criterion is chosen to be a global one, because there is a limit on the convergence that can be achieved, which is not known a priori. Namely for a given $\mathcal{P}(\delta_{ng})$ and $P(k)$, $\epsilon$ cannot be made arbitrarily small. This happens because the PDF and the correlation function may not be compatible. There are two kinds of incompatibilities \citep{Vio01}: one arises when the target correlation has values that lie outside the range that can be reached by the correlation function corresponding to the chosen translation transform $g$. The other incompatibility arises when the target correlation is not positive definite, which is a necessary requirement for a well defined correlation function. The updating scheme \cref{eq:update} is such that a strictly positive target power spectrum must correspond to a strictly positive pre-translation spectrum. Clearly this holds true also when a Gaussian smoothing is adopted, so that the result of ITAM is always a well defined correlation funtion. Even in the case when the first kind of incompatibility is present, ITAM still provides the best one could do if the goal is to make a fast simulation with a translation process (see \cite{Shields11} for some examples).

The speed and accuracy of the convergence depend on $\beta$, which reflects the change in the updated Gaussian power. Too-high values for $\beta$ could result in an overshoot with respect to the target. Likewise, values too small could render the updating ineffective. We find that is better to choose values such as $\beta \sim 0.2 - 1.5$ and see how small $\epsilon$ can be made, and finally select the one which delivers the smallest value. This allows to select $\alpha$ so that ITAM stops when $\epsilon<\alpha$. Alternatively, ITAM can stop when the relative change $\Delta \epsilon/\epsilon$ between consecutive iterations becomes small. This is independent of how small $\epsilon$ is, and reflects more closely the fact that the algorithm converged as well as it could to the target. This fine-tuning procedure needs to be performed only once.

Ultimately, the output of ITAM is the power spectrum $P_g(k)$ that allows realizations of pre-translation Gaussian fields to be transformed into the non-Gaussian ones by means of the translation transform \cref{eq:g}. ITAM can be applied to any stationary random field of interest in a three dimensional box; we provide the code publicly\footnote{\url{https://github.com/tos-1/ITAM}}.

\section{Application to the Matter Density field}\label{sec:Results}

\subsection{Method}
As a demonstration, we now investigate the use of ITAM in the cosmological case of the matter density field. The target power spectrum can be generated using one of the many useful codes available in the literature. We choose to generate the target nonlinear power spectrum with \textsc{HALOFIT} \citep{Takahashi12}, implemented in \textsc{CLASS} \citep{Blas11}.

As a next step, we need a template PDF for our target field. The task of providing a comprehensive analytical approximation to the PDF of the density field is a difficult one. \cite{Klypin17} shows how the PDF depends on the simulation box size, resolution, redshift, smoothing scale and cosmological parameters. In \cite{Uhlemann16} an analytical derivation of the PDF for the quasi nonlinear scales is presented, and in \cite{Uhlemann19} its complementarity with the power spectrum on cosmological parameter constraints is shown. Both \cite{Uhlemann16} and \cite{Shin17_PDF} argue that the PDF deviates significantly from the lognormal distribution, and it is better approximated by a skewed lognormal distribution. This result is confirmed by \cite{Repp18}, who show that a Generalized Extreme Value distribution is a better fit to the matter field PDF when transitioning to nonlinear smoothing scales. Despite these efforts, there is still no general scheme that provides a fitting function, implemented in a publicly available software and with arbitrary resolution, so we prefer to use the exact PDF measured from simulations. While this could bias the results towards a specific realization rather than a template average, it is still good enough as a proof of concept.

We take two simulations as references. The first is the Millenium run (ML) \citep{Millenium}, a simulation of size $500$ \hmpc, with $2160^3$ particles, and with a WMAP cosmology $h=0.73$, $\Omega_b = 0.045$, $\Omega_m = 0.205$, $\Omega_{\Lambda}=0.75$, $n=1$ and $\sigma_{8}=0.9$. We use a $256^3$ nearest-grid-point density grid, provided by the ML database. We also consider an $N$-body simulation with a smaller box size (hence SB), of size $256^3$\hmpc\ and $256^3$ particles. The simulation was run with Gadget \citep{Gadget}, starting at an initial redshift of $z=50$, and with initial conditions set by second-order Lagrangian perturbation theory (2LPT). The fiducial cosmology assumed is a vanilla $\Lambda$CDM cosmological model, with $h=0.7$, $\Omega_b = 0.046$, $\Omega_{\rm CDM}=0.25$, and $\sigma_{8}=0.8$. The amplitude of the modes of the initial conditions is set to be precisely the ensemble-mean power spectrum, following the suggestion of \cite{Angulo16}, to have a better convergence on the ergodic properties. While fixing the variance does not improve the convergence on the ensemble-mean PDF, it was shown not to bias it for smoothing scales as small as $0.5$ \hmpc\ \citep{Klypin20}. For both simulations, we sample the matter field by the grid resolution, which is $R_s = 1.95 \ h^{-1} Mpc$, $R_s = 1.0 \ h^{-1} Mpc$ for the ML and SB simulations, respectively. The smoothing is performed with a Gaussian kernel
\begin{equation}\label{eq:smoothing1}
	\delta(k,\sigma_s) = \delta(k) e^{\frac{-k^2 R_s^2}{2}},
\end{equation}
and consequently the target power spectrum has to be smoothed as well
\begin{equation}\label{eq:smoothing2}
	P(k,\sigma_s) = P(k) e^{-k^2 R_s^2}.
\end{equation}

For the algorithm to converge accurately, it is important to test beforehand that the variance of the PDFs is consistent with that computed from their target power spectra, as this could result in an offset of the normalization of the power spectrum. The variance of the PDFs can be directly computed as the cell density variance of \cref{eq:smoothing1} in real space, while the variance of the correlation function is simply \cref{eq:WK} computed at $r=0$, with the addition of the smoothing Gaussian kernel in the integration. In case there is an offset, the correlation function should be rescaled to make them consistent.

The power spectrum must be smoothed, to limit the variance of the field; without smoothing, \cref{eq:WK} would diverge when evaluated at $r=0$. Alternatively, the smoothing can be avoided by considering the PDF of the field sampled at the grid resolution. In this case, there is an implicit filtering to be considered, i.e. a sharp cut-off of the target power spectrum at the Nyquist mode of the box. In the following section we will show how ITAM performs in both cases.
\begin{figure}
	\includegraphics[width=0.45\textwidth]{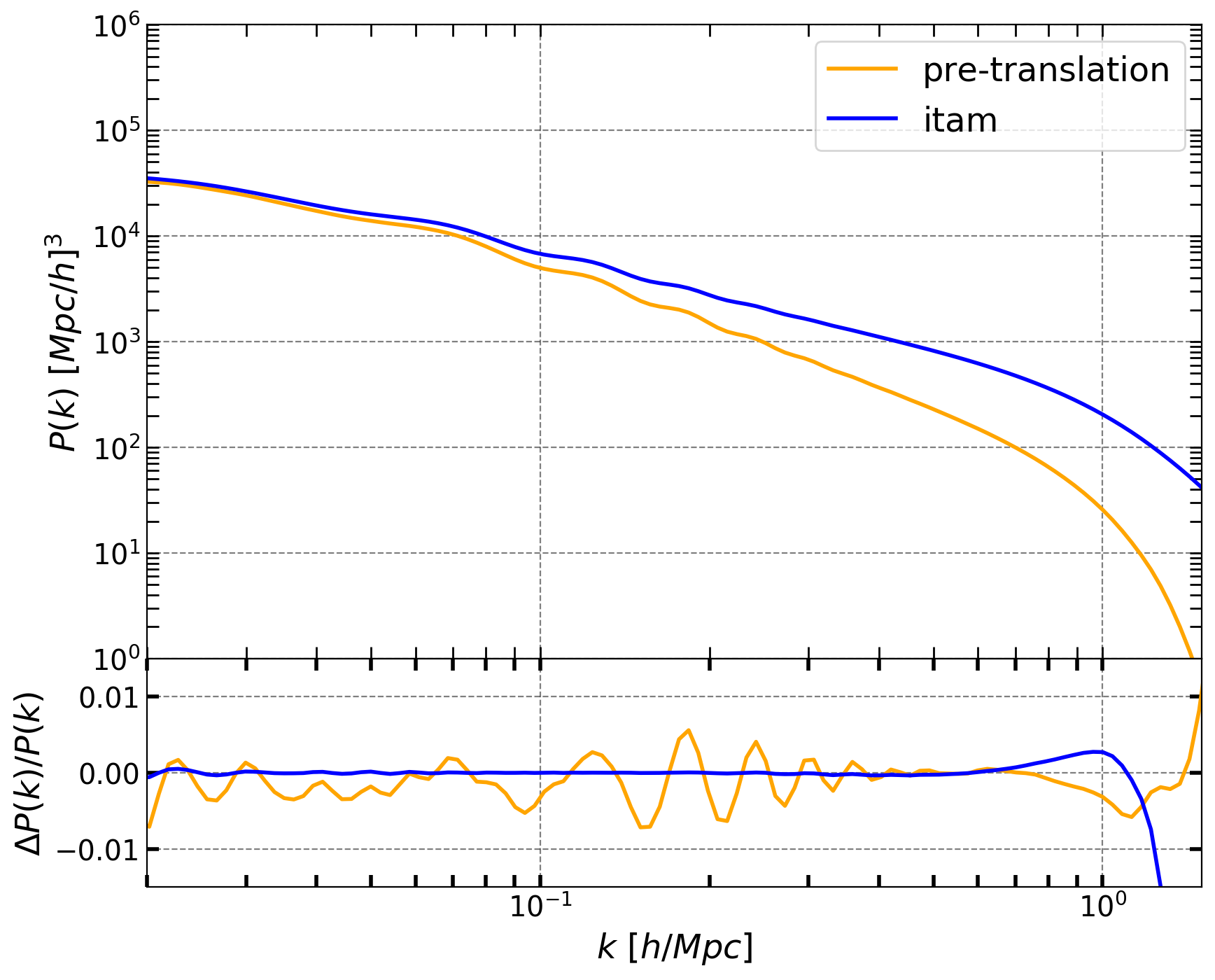}
	\caption{ Results of ITAM for the mapping \cref{eq:CJ}. The top panel shows the power spectrum achieved by ITAM, while in the bottom its relative difference with respect to the target is shown (blue line). The pre-translation Gaussian spectrum that results from the optimization process is also shown, and its relative difference with respect to the Gaussian one predicted by \cref{eq:ln2} is plotted in the bottom (yellow line). The smoothing scale considered is 1 \hmpc. \label{fig:CJ_rel_diff_spectra} }
\end{figure}

\begin{figure}
	\includegraphics[width=0.45\textwidth]{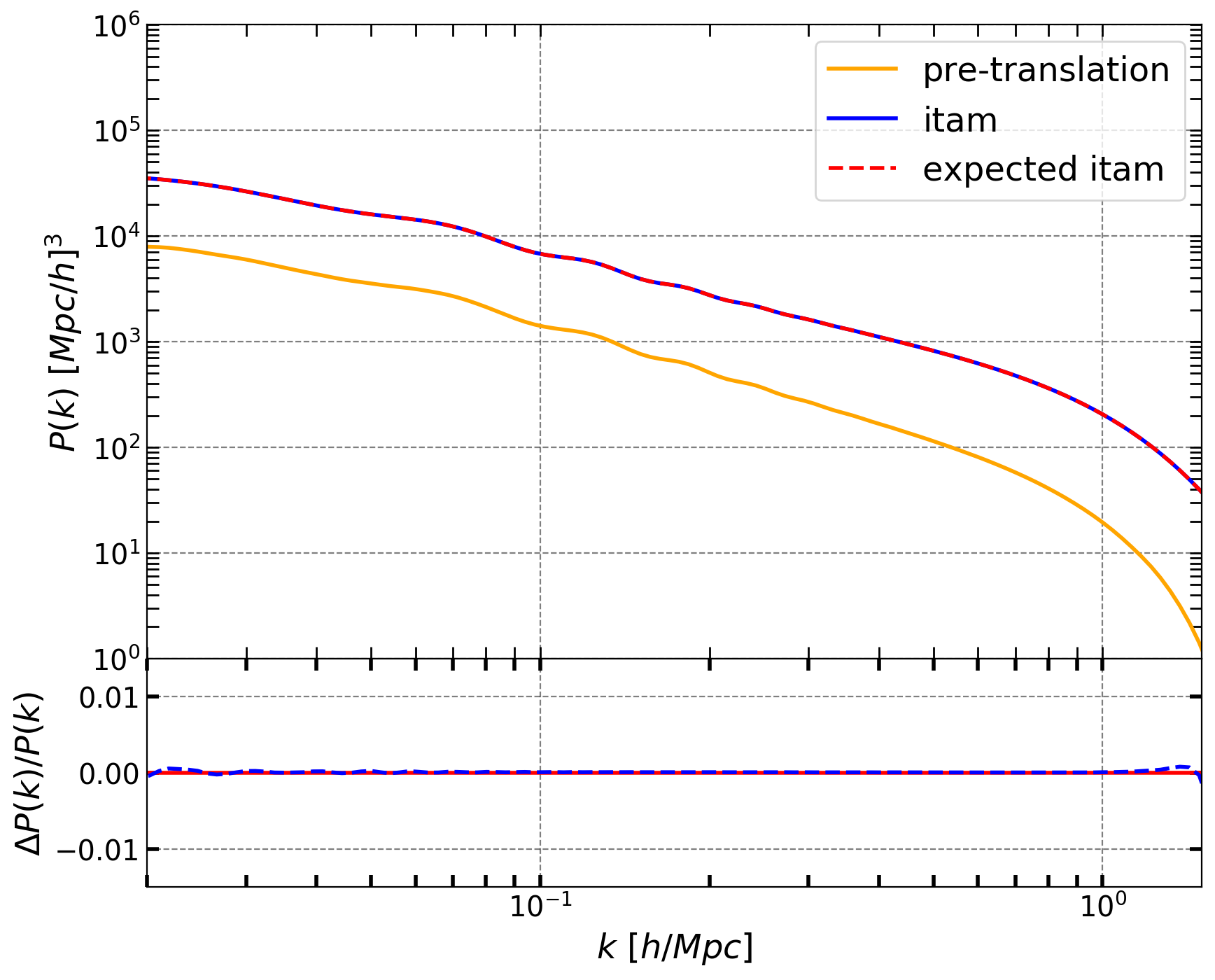}
	\caption{ Results of ITAM for the mapping \cref{eq:CJ_offset}. The top panel shows the nonlinear power spectrum that can be achieved with ITAM (see legend). In the bottom it is shown the relative difference with respect to the target (dashed blue line). The \textit{expected itam} curve shows the theoretical prediction for the nonlinear spectrum computed from \cref{eq:ln4}, with its relative difference with respect to \textit{itam} shown in the bottom panel (solid red line). The pre-translation spectrum is offset with respect to the translated one (\cref{eq:ln4}). The smoothing scale considered is 1 \hmpc. \label{fig:log_rel_diff_spectra}}
\end{figure}

\subsection{Results}
Before applying ITAM to the case of the exact PDF measured from $N$-body simulations, we apply it by assuming a lognormal PDF for the target field, for which we can derive the exact solution of the inverse problem. To run ITAM we first initialize a pre-translation Gaussian power spectrum. The choice is not relevant as long as it is a sensible one -- e.g. one could choose the very same target spectrum or the linear density power spectrum for the initialization. Following the scheme outlined in \cref{fig:FlowChart}, we transform this power spectrum in real space to compute \cref{eq:Grigoriu}, for which we need to specify the translation transform \cref{eq:g}, namely the combination of the CDF of a zero-mean Gaussian variable and the inverse CDF for the associated lognormal
\begin{equation}\label{eq:CJ}
	\delta^L_{ng} = e^{-\sigma^2/2} \mathcal{F}_L^{-1}\left[\mathcal{F}_g[\delta_g]\right]-1.
\end{equation}

The prefactor is necessary to adjust the inverse CDF of the lognormal to map onto a zero mean field. This mapping effectively corresponds to \cref{eq:ln}, as we show in \cref{fig:CJ_rel_diff_spectra}.   
As a separate test, we implement a transform where we ignore the prefactor $e^{-\sigma^2/2}$, so that the mapped field is a lognormal field with nonzero mean. We can compute the expected mean $\mu$ that must be subtracted in \cref{eq:Grigoriu} by considering that in this case the lognormal transform is
\begin{equation}
	\delta^L_{ng} +1 = e^{\delta_g},
\end{equation}
whose expected mean is
\begin{equation}
	\mu = e^{\sigma^2/2} - 1,
\end{equation}
so that the corresponding mapping to a zero mean lognormal field is
\begin{equation}\label{eq:ln3}
	\delta^L_{ng} = e^{\delta_g} - e^{\sigma^2/2}.
\end{equation}   

\begin{figure}
	\includegraphics[width=0.43\textwidth]{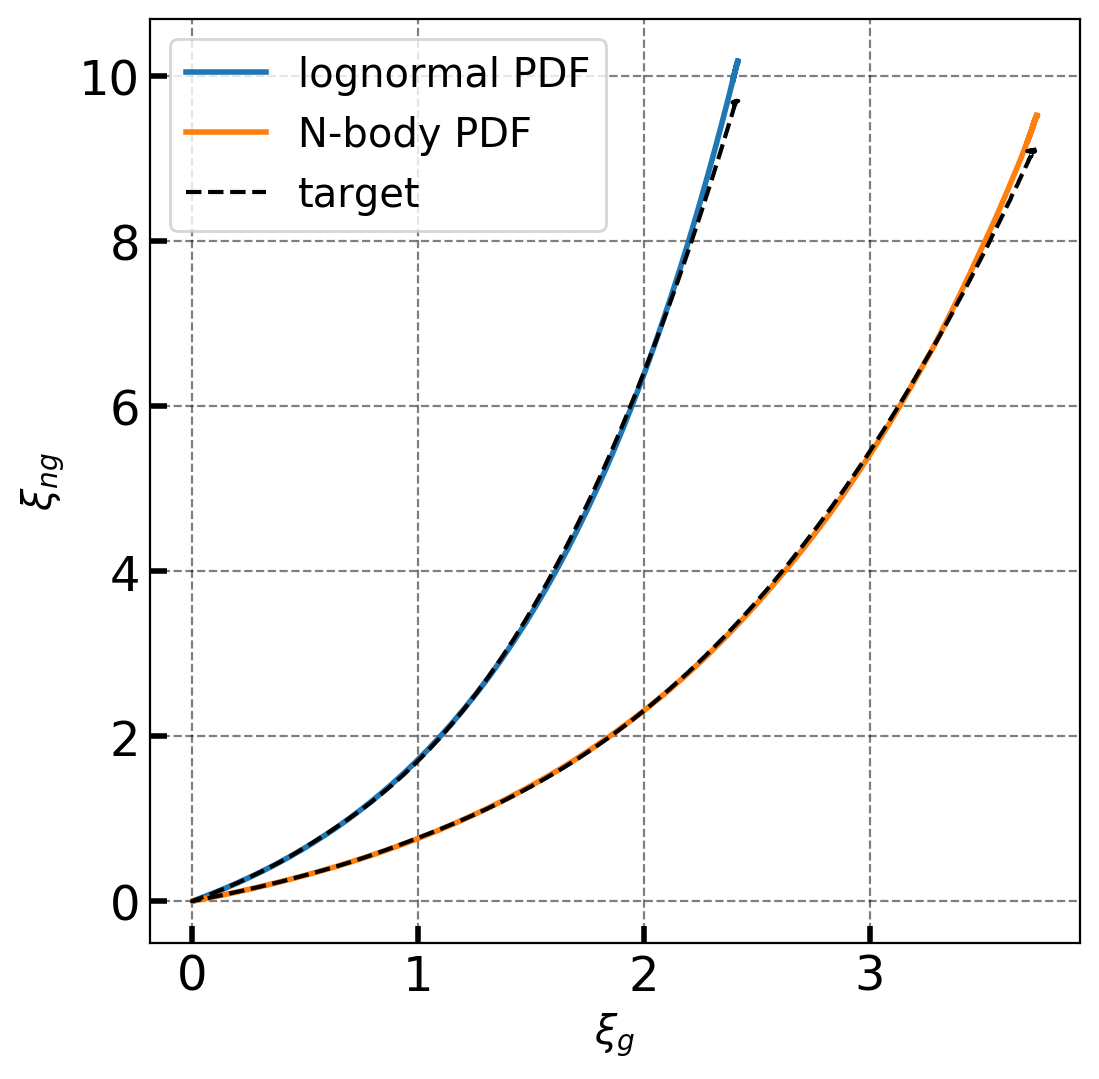}
	\caption{Results of the optimization process of ITAM in real space. The target power spectrum and the N-body PDF from the SB simulation were both smoothed at 1 \hmpc. The result for the lognormal PDF was also obtained by running ITAM with the transform \cref{eq:CJ}. At this smoothing scale, we can see that a linear relation $\xi_{ng}(r) \propto \xi_{g}(r)$ holds for $\xi_g \lesssim 1$, which corresponds to $r\gtrsim6$ \hmpc. \label{fig:correlations}}
\end{figure}

\begin{figure}
	\includegraphics[width=0.45\textwidth]{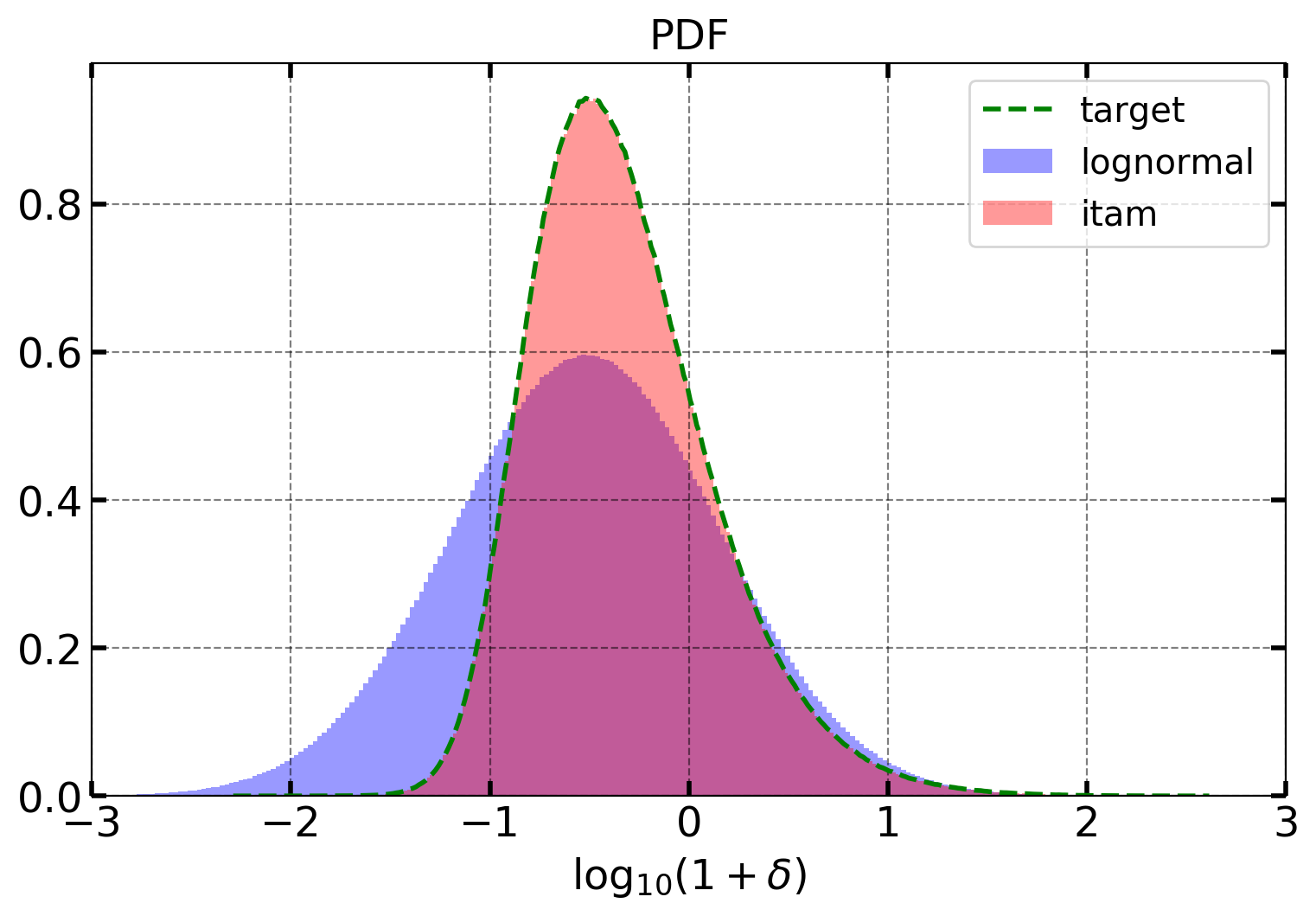}
	\caption{PDF of an ITAM density field realization compared to a lognormal density field realization, and to the expected target PDF (dashed line). The smoothing scale for both realizations is 1 \hmpc.\label{fig:PDFs}}
\end{figure}

\begin{figure}
	\includegraphics[width=0.45\textwidth]{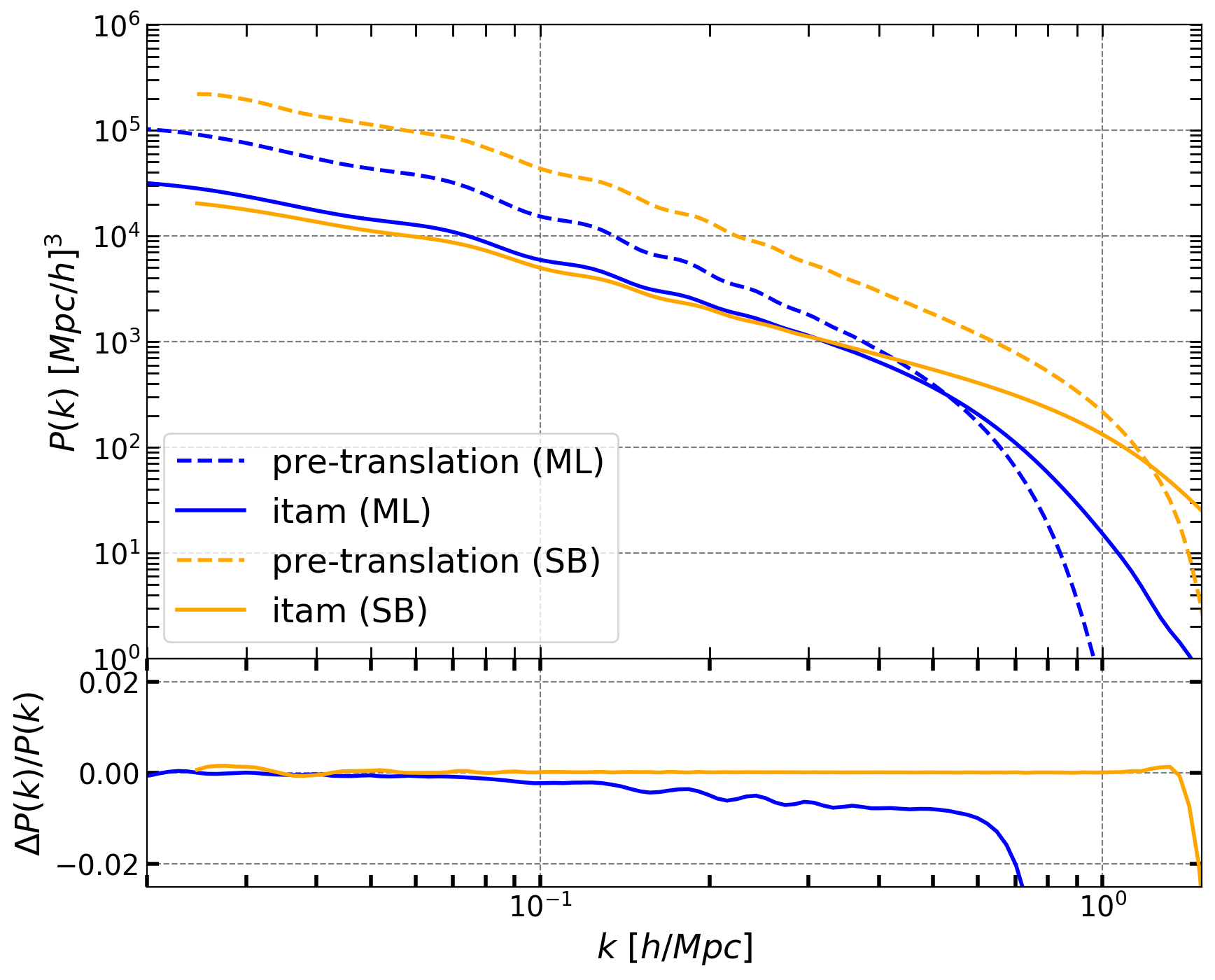}
	\caption{The top panel shows the pre-translation spectra and the nonlinear spectra output from ITAM, for both the simulations. The bottom panel shows the relative difference with respect to the target nonlinear spectra. The smoothing scale introduces a cut-off that reduces the convergence on smaller scales. Also, the increased smoothing scale adopted in the ML simulation (2 \hmpc) with respect to the SB simulation (1 \hmpc) manifests itself as a smaller offset between the pre-translation spectrum and the target spectrum, thus showing that it is due to the skewness of the distribution. \label{fig:Gauss_rel_diff_spectra}}
\end{figure}

\begin{figure}
	\includegraphics[width=0.45\textwidth]{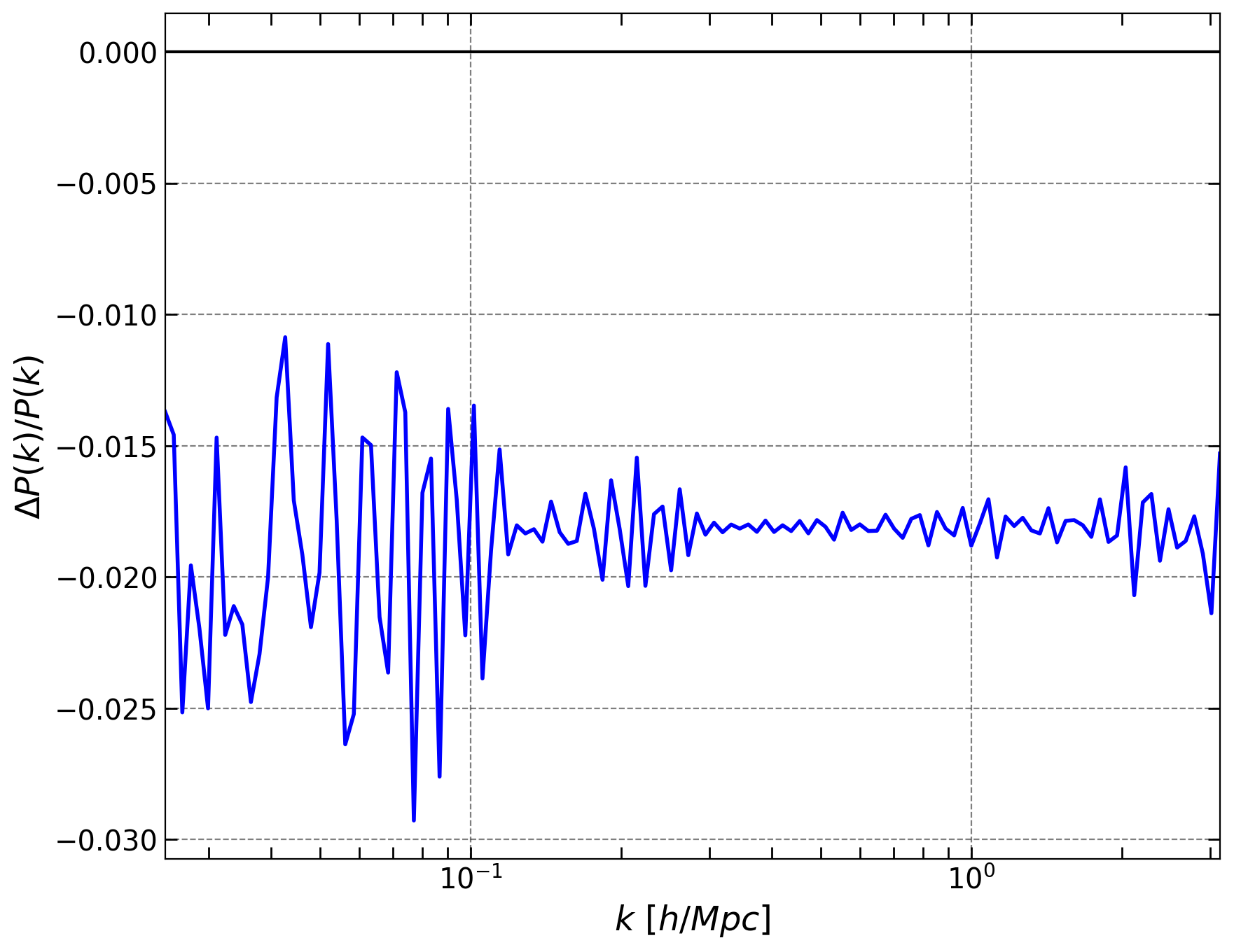}
	\caption{Residuals of the ITAM power spectrum with respect to the target, in the case of the PDF from the SB simulation. The field has not been filtered, while a sharp cut-off for the target power spectrum in correspondence with the Nyquist mode of the box is adopted. While in the unsmoothed case ITAM is able to recover the signal better up to the Nyquist mode, the global convergence gets worse with respect to the Gaussian smoothing case. \label{fig:shapcoff}}
\end{figure}

\begin{figure}
	\includegraphics[width=0.45\textwidth]{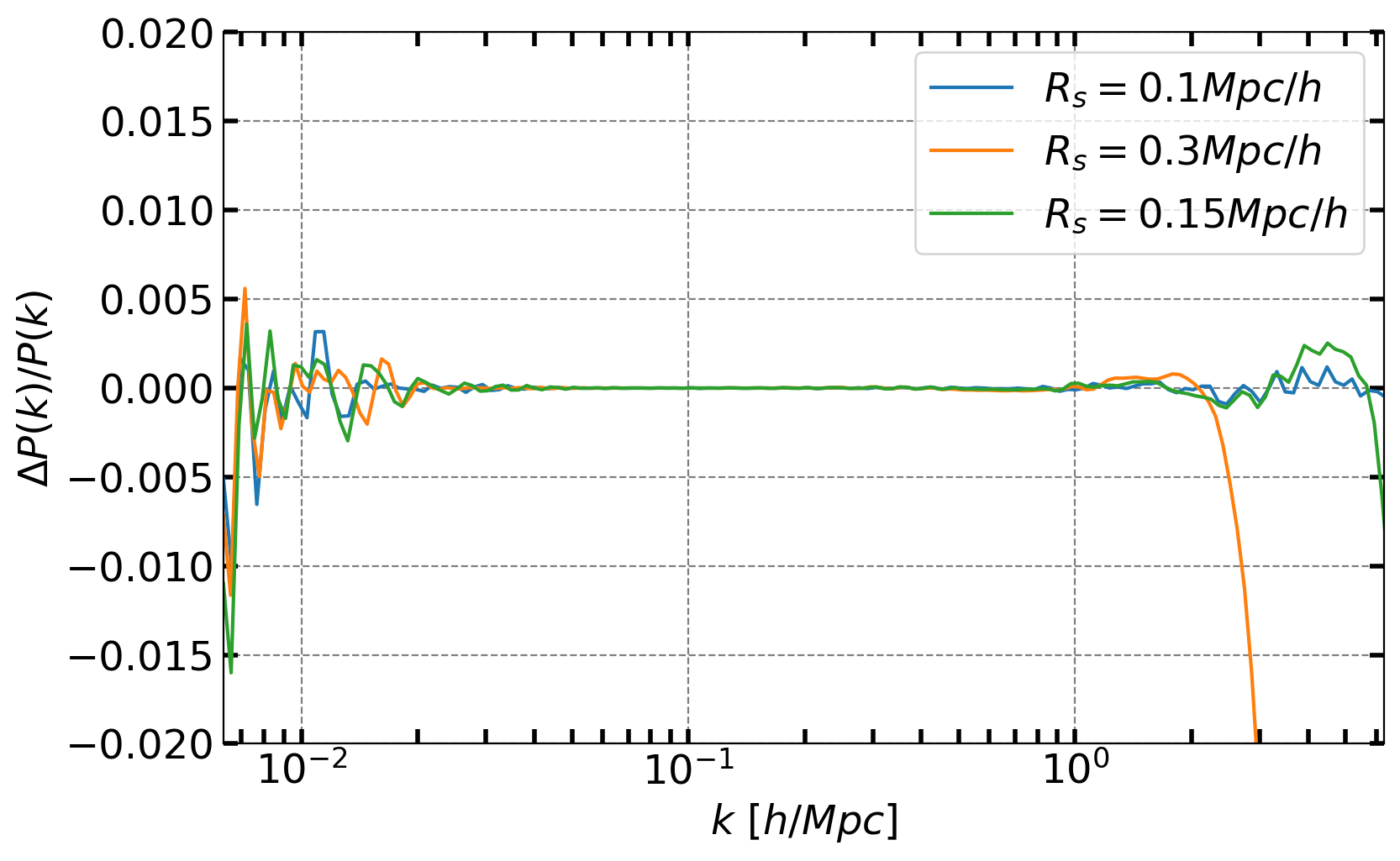}
	\caption{Tests of ITAM in the case of a lognormal target PDF for a simulation of box size 1000 \hmpc, by varying the smoothing scale $R_s$. \label{fig:rel_resolution}}
	%\FT{The sampling rate per side $n_g$ is not directly used in the optimization process of ITAM, but it is just shown as a reference of the sampling rate that can be chosen to make simulations accurate up to the Nyquist frequency. In all the cases the smoothing scale satisfies the condition to be smaller than the sampling rate, as suggested in \citep{Baratta20}.}\label{fig:rel_resolution}}
\end{figure}

In analogy to \cref{eq:CJ}, the translation transform we implement in this case is
\begin{equation}\label{eq:CJ_offset}
	\delta^L_{ng} +1 = \mathcal{F}_L^{-1}\left[\mathcal{F}_g[\delta_g]\right]-e^{\sigma^2/2}+1,
\end{equation}
which is still set by matching the CDF of the Gaussian to the CDF of the lognormal, knowing that the expected mean is $\mu=e^{\sigma^2/2}-1$. While this lognormal field does not correspond to a density field, having $\delta^{L}_{ng}<-1$, it is still useful to consider, as we can use \cref{eq:ln3} to compute the expected correlation function to make a comparison with the numerical results of ITAM. The expected correlation function is
\begin{equation}\label{eq:ln4}
	\xi^L_{ng}(r) = e^{\sigma^2} (e^{\xi_{g}(r)} -1)
\end{equation}
(see \cite{Xavier16} for a derivation of correlations for general lognormal fields). In this case we cannot solve explicitly for the pre-translation Gaussian correlation, as also the prefactor depends on the Gaussian correlation via $\sigma^2=\xi_g(0)$. However, we can still check whether the optimized pre-translation correlation function satisfies \cref{eq:ln4}; this is shown in \cref{fig:log_rel_diff_spectra}. It is worth noticing that in both cases examined above, $\sigma$ varies during the optimization process as well, as it is the standard deviation of the pre-translation field.\\
\indent We proceed to examine the results of ITAM when the input PDF is measured from the SB and ML simulations. We already know that the nonlinear density field has zero mean, so that in order to fully specify the translation transform \cref{eq:g}, it is enough to use the Gaussian CDF together with a lookup table of the inverse CDF of the smoothed target field. ITAM can be run after both the initialization and target power spectra have been chosen. The result of the optimization can be seen in real space in \cref{fig:correlations}, which shows that it is equivalent to solve $\xi_{ng}$ from \cref{eq:Grigoriu} as a function of $\xi_g$. An example of the PDF of a density field generated with ITAM is shown in \cref{fig:PDFs}, compared with the approximation of a lognormal realization.\\
\indent
In \cref{fig:Gauss_rel_diff_spectra} we compare the output power spectrum that can be obtained by ITAM relative to the target. We can see that for both PDFs measured from our two simulations, SB and ML, percent accuracy is reached up to half of the Nyquist mode of the box, with a more precise result for the SB simulation. This is due to the fact that the convergence to the target spectrum is affected by the smoothing scale. The steep cutoff introduced by the smoothing can be seen in the drop of the residuals plotted at the bottom of \cref{fig:Gauss_rel_diff_spectra}. In the case the PDF is measured from a simulation, the choice of the smoothing scale cannot be arbitrary, but depends on the mass assignment scheme used to interpolate the density on the grid, and on the sampling rate of the simulation. We find that smoothing at the scale of the sampling rate of the simulation guarantees the best convergence for the optimization.
In the case we use the PDF of the field sampled at the grid resolution, the overall convergence becomes less accurate, but the target power spectrum is recovered with increased precision up to the Nyquist mode, as shown in \cref{fig:shapcoff}. It seems that there is an interplay between large scale and small-scale power: the larger the small scale power, the more difficult the global convergence to the target, perhaps because a PDF with a larger tail is more difficult numerically. For the sake of precision cosmology, we think the Gaussian smoothing is preferable: one can always employ the PDF from a high-resolution simulation to compensate for the loss of signal at smaller scales, and eventually downsample the grid to reach a better accuracy up to the Nyquist mode for the fast realizations.\\
\indent To prove that the Gaussian filtering does indeed work for smaller smoothing scales than those considered until now, we run ITAM in the case of the lognormal transform \cref{eq:CJ} examined before. In this case we are not limited by the accuracy of the sampled PDF, as it is analytical, and we can adopt arbitrary smoothing scales, shown in \cref{fig:rel_resolution}. This confirms that ITAM can in principle be pushed to smaller scales than the ones we examined with finer grids and higher-resolution PDFs. ITAM is independent of the resolution, which means that its output power spectrum can be used to make realizations at the desired sampling rate, so that is always possible to adjust the grid in order to recover a realization accurate up to the Nyquist mode.\\
\indent From \cref{fig:Gauss_rel_diff_spectra} we notice that the pre-translation Gaussian spectrum output from ITAM has an offset with respect to the target power spectrum, even on large scales. This offset can also be seen in \cref{fig:correlations} in the Gaussian correlation, and it is due to the fact that we did not add a scaling factor, which is present in the case of \cref{eq:CJ}. In fact, for the actual PDF of the density field we ignore what is the correct rescaling, but one can see that it is related to the skewness of the distribution. While in the lognormal case the skewness is set by specifying the variance of the underlying Gaussian field, distributions more skewed than the lognormal require additional parameters to characterize them \citep{Shin17_PDF,Repp17}, and they break this one to one correspondence. From \cref{fig:Gauss_rel_diff_spectra} one can see that a more skewed distribution results in an increased offset, namely a scaling factor that has to be smaller than $e^{-\sigma^2/2}$ of \cref{eq:CJ}. Ultimately, this rescaling is only convenient for visualization purposes, but it is not relevant for the aim of generating nonlinear fields, as ITAM can correctly take it into account in the optimization process. We caution that these pre-translation fields should not be taken to be physically meaningful for cosmological purposes.

\subsection{Covariance Matrix from ITAM}
Having optimized for the linear power spectrum, we can generate many samples of Gaussian fields to transform into the target density field. The skewness and the kurtosis of the PDF are related to integrals of the bispectrum and the trispectrum, so we expect the ITAM scheme to provide a small improvement over the standard lognormal case for the corresponding higher-order polyspectra. We test this hypothesis at the level of the covariance matrix of the power spectrum, defined as
\begin{equation}
	C_{ij} = \frac{1}{N-1} \sum^{N}_{n=1} \left( P(k_i) - \overline{P}(k_i) \right) \left( P(k_j) - \overline{P}(k_j) \right),
\end{equation}
where $N$ is the number of realizations, and $\overline{P}(k_i)$ is their average power spectrum, estimated for each simulation as 
\begin{equation}
	\overline{P}(k) = \frac{1}{N_k} \sum_{|\pmb{k}| \in k} |\delta(\pmb{k})|^2,
\end{equation}
where $N_{k}$ is the number of independent modes in the sum. For the PDF, we use the ML simulation, as it has a bigger volume. To validate our covariance matrix, we choose as a theoretical model the phenomenological description found by \cite{Neyrinck11_cov}. There, the matter power-spectrum covariance from $N$-body simulations was found to be accurately approximated by
\begin{equation}\label{eq:covariance}
	C_{ij} = \delta_{ij} 2 \frac{P^2(k_i)}{N_{k_i}} + \alpha P(k_i) P(k_j),
\end{equation}
hence we refer to this model as $\alpha$ \textit{model}. This model was later revisited by \cite{Mohammed14,Carron15}. In \cite{Neyrinck11_cov} the $\alpha$ parameter was fitted against the covariance of a suite of $N$-body from the Coyote Universe simulations \citep{Lawrence10}, but its value could also be estimated directly from the cell density variance of each realization as $\alpha={\rm Var}(\sigma^2_{\rm cell})/\langle\sigma^2_{\rm cell}\rangle^2$, with $\sigma^2_{\rm cell}$ being the cell density variance of the field, and $\langle\sigma^2_{\rm cell}\rangle$ the corresponding ensemble average over the realizations. In order to make a comparison we generate $5000$ realizations, both using ITAM and lognormal approximation, and we compute their power spectra, their covariance and the corresponding $\alpha$ parameters.

\begin{figure*}
	\includegraphics[width=0.95\textwidth]{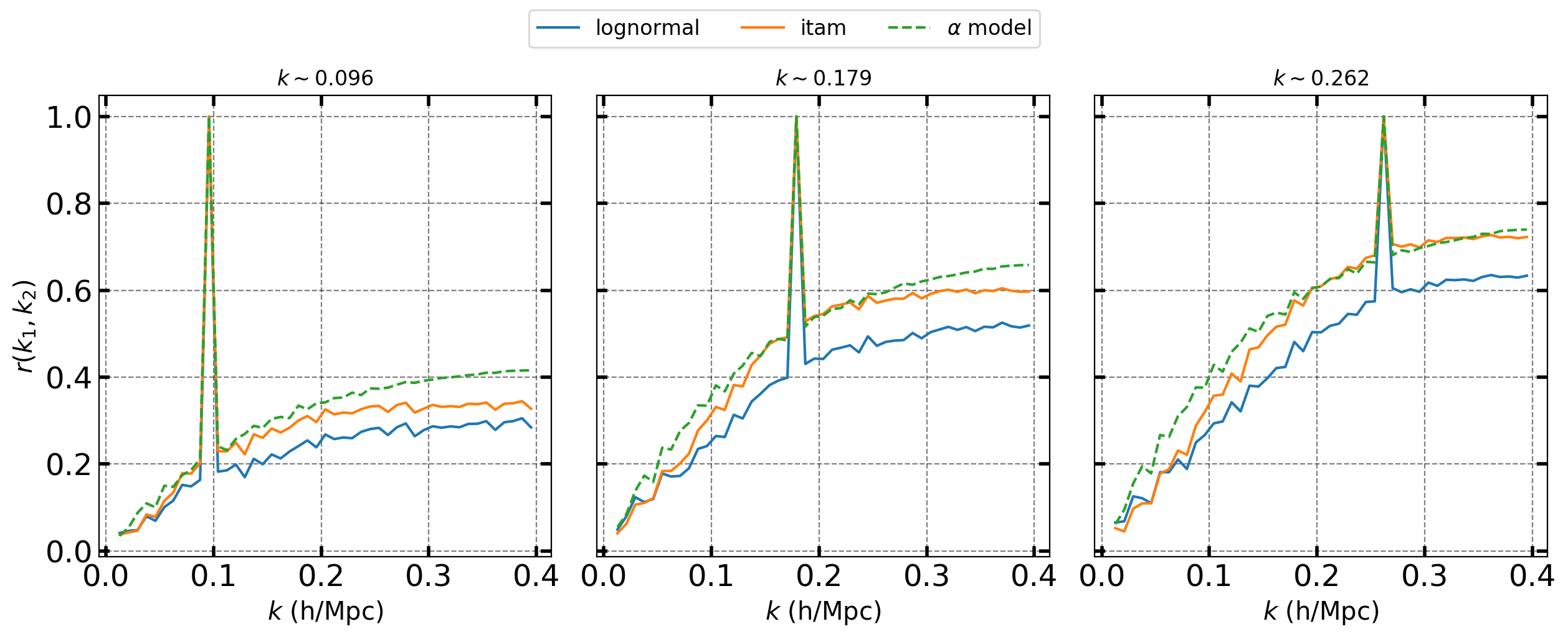}\caption{ Cut through the correlation matrix for the three approximations examined; the fixed mode is indicated at the top of each panel. \label{fig:off_diagonal}}
\end{figure*}

\begin{figure}
	\includegraphics[width=0.45\textwidth]{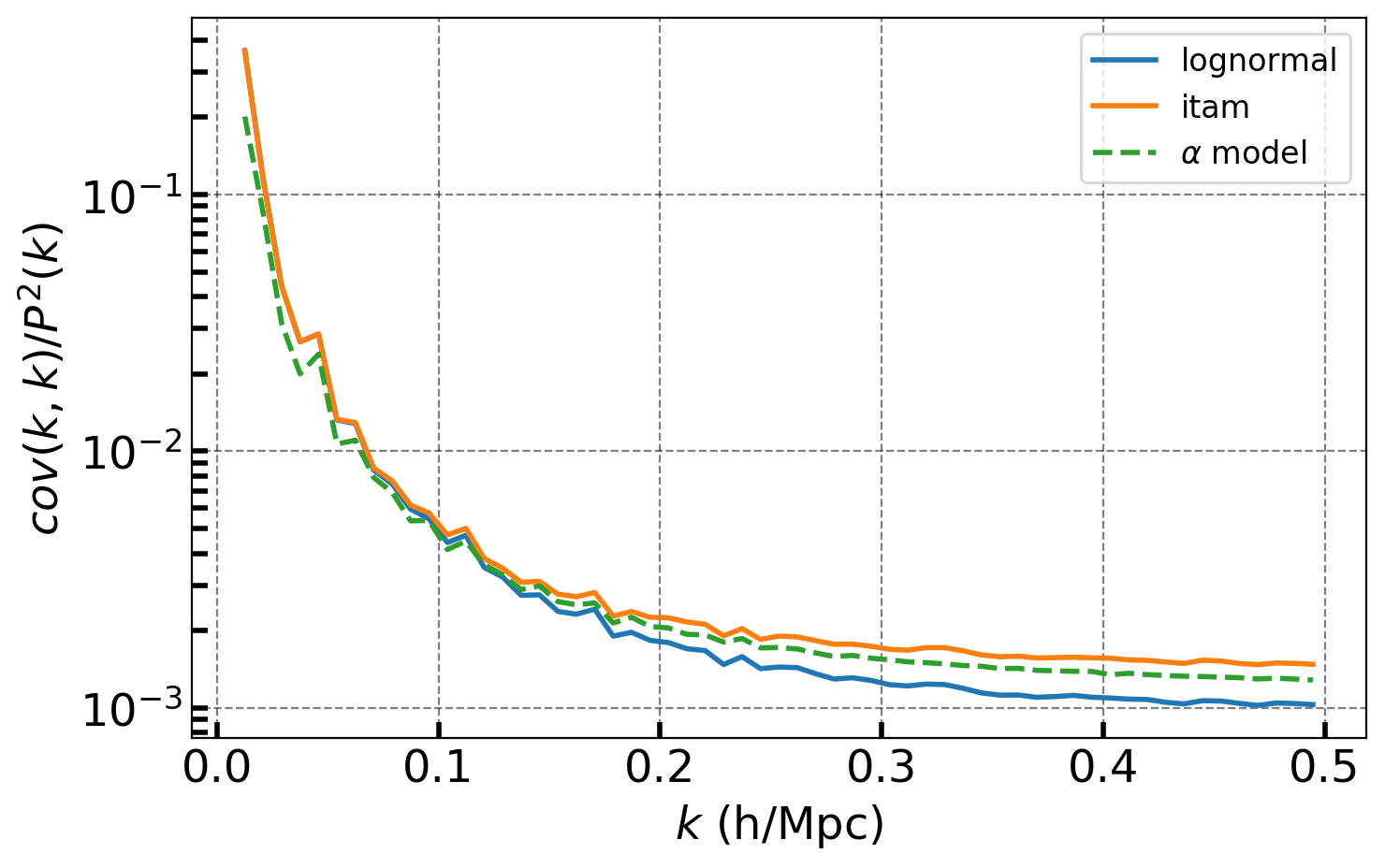}\caption{
		Diagonal elements of the covariance matrix for the three approximations examined in the text. \label{fig:diagonal}}
\end{figure}

We compare the structure of the covariance matrices by analyzing both the off-diagonal (\cref{fig:off_diagonal}) and diagonal elements (\cref{fig:diagonal}). We can see that either cases retain a similar shape to the $\alpha$ model, with a minor improvement of the ITAM case over the lognormal one. In \cref{fig:Tij}, both the lognormal and ITAM follow rather well the $N$-body fit provided by the $\alpha$ model, shown in terms of the non-Gaussian contribution to the covariance. This is confirmed by the $\alpha$ values that we measured for either approximation schemes in two ways, both by employing $\alpha={\rm Var}(\sigma^2_{\rm cell})/\langle\sigma^2_{\rm cell}\rangle^2$ at the level of sample realizations, and by fitting the functional form \cref{eq:covariance} to the covariance matrices. In the case of the lognormal, we find $0.0009$ for the former and $0.0014$ for the latter method, while in the case of the ITAM scheme we have $0.0013$ and $0.0014$ respectively, to be compared against the reference value $0.0011$ found by fitting \cref{eq:covariance} to the covariance of the Coyote Universe simulations \citep{Lawrence10} used by \cite{Neyrinck11_cov}.

\subsection{Relation to multiplicative bias model}\label{subsec:bias}
\cite{Neyrinck11_cov,Carron15} show that the approximation to the covariance provided by \cref{eq:covariance} is exact for a $b \delta$ field, where $\delta$ is a Gaussian field with power spectrum $P(k)$ and $b$ is a multiplicative `bias' field, i.e.\ a locally constant factor multiplying the density field, which varies from patch to patch, such that $\langle b^2 \rangle=1$ and $\langle b^4 \rangle-\langle b^2 \rangle^2= \alpha$.
%In other words, the $\alpha$ model covariance is exact for a Gaussian field that undergoes an amplitude modulation on scales much larger than those probed by the power spectrum.

\begin{figure*}
	\includegraphics[width=0.95\textwidth]{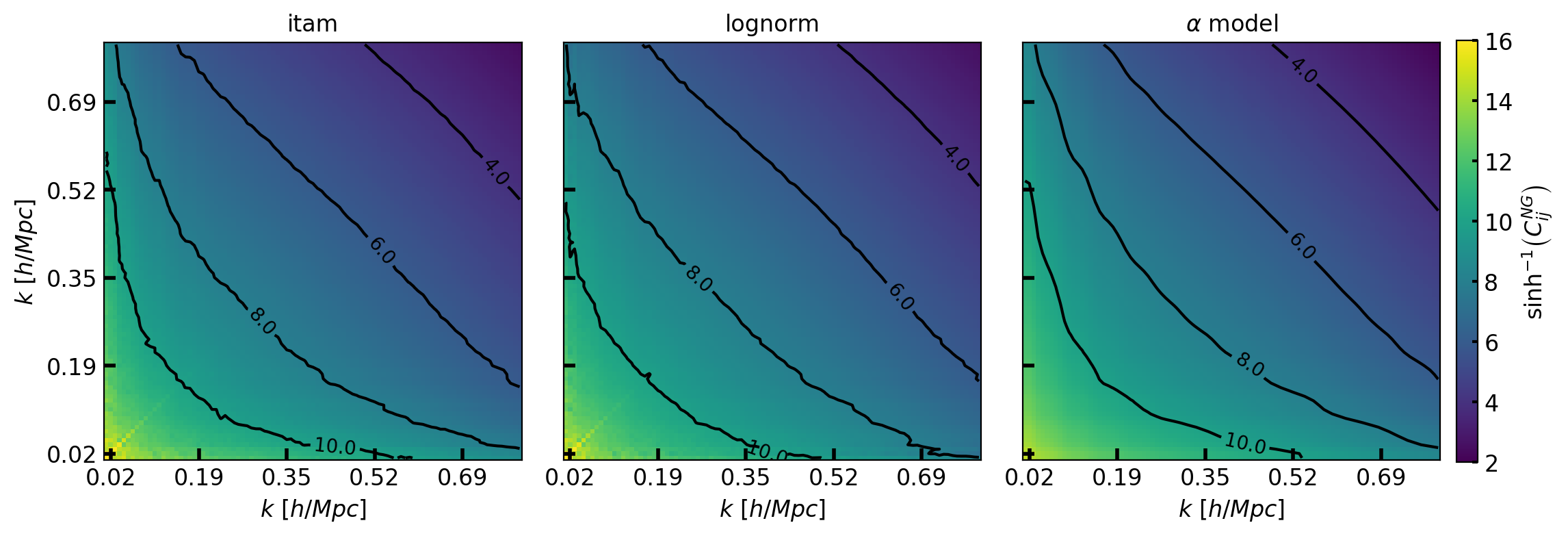}\caption{ Non-Gaussian part of the covariance matrix, $C^{NG}_{ij} = C_{ij}-2\delta_{ij}P^2(k_i)/N_{k_i}$. The black lines denote the isocontours. \label{fig:Tij}}
\end{figure*}

The fact that translation processes (ITAM, and the lognormal models) approximate the covariance in \cref{eq:covariance} suggests that the bias fluctuations can be ascribed to the monotonic mapping of the initial density fluctuations. The non-Gaussian components of ITAM and lognormal covariances retain a very similar structure to the $\alpha$ model approximation, as shown in \cref{fig:Tij}. Moreover, the covariance structure of an $N$-body density field can be largely removed by Gaussianizing or taking the log-density of the field (e.g. fig. 2 of \cite{Neyrinck11_cov}). In the case of the lognormal field, the exact computation can be performed, and we can attempt to see whether the form \cref{eq:covariance} is an approximation for its covariance.

We already considered the relation between lognormal and Gaussian fields (\cref{eq:ln}) in terms of their correlation functions (\cref{eq:ln2}). It is possible to generalize the relation between the correlation functions of the two fields to arbitrary orders \citep{Coles91}:
\begin{equation}\label{eq:highcorr}
\begin{split}
	\langle (\delta^L_{ng}(\pmb{x}_1) + 1)&...(\delta^L_{ng}(\pmb{x}_n) + 1) \rangle =\\
	&\exp{\left( \sum_{i<j}^n \xi_g(|\pmb{x}_i-\pmb{x}_j|) \right)} = \prod_{i<j}^n \left[1+\xi(|\pmb{x}_i-\pmb{x}_j|) \right],
\end{split}
\end{equation}
where $\xi$ is now used to refer to the two-point correlation function of the lognormal field. We also know that the three-point correlation of the lognormal field can be written as
\begin{equation}\label{eq:3point}
\begin{split}
	\langle (\delta^L_{ng}(\pmb{x}) + 1)&(\delta^L_{ng}(\pmb{x}+\pmb{r}_1) + 1)(\delta^L_{ng}(\pmb{x}+\pmb{r}_2) + 1)) \rangle=\\
	&\Gamma(\pmb{r}_1,\pmb{r}_2) + \xi(r_1)+ \xi(r_2)+ \xi(r_{12})+1,
\end{split}		
\end{equation}
where $\Gamma$ is the connected three-point correlation function. By combining \cref{eq:highcorr} and \cref{eq:3point}, we can solve for the connected component
\begin{equation}\label{eq:connected3point}
\begin{split}
	\Gamma(\pmb{r}_1,\pmb{r}_2)&= \xi(r_1)\xi(r_2)+\xi(r_1)\xi(r_{12})+\xi(r_2)\xi(r_{12})\\&+\xi(r_{1})\xi(r_{2})\xi(r_{12}).
\end{split}
\end{equation}

In the same way we compute the fourth order correlation, which depends not only on the two-point correlation function, but also on the three-point connected component
\begin{equation}\label{eq:4point}
\begin{split}
\langle(\delta^L_{ng}(\pmb{x}) &+ 1)(\delta^L_{ng}(\pmb{x}+\pmb{r}_1) + 1)(\delta^L_{ng}(\pmb{x}+\pmb{r}_2) + 1))(\delta^L_{ng}(\pmb{x}+\pmb{r}_3) + 1)) \rangle\\&=
	1 + \xi(r_{1}) + \xi(r_{21}) + \xi(r_{2}) + \xi(r_{31}) + \xi(r_{32}) + \xi(r_{3})\\ &+ \xi(r_1)\xi(r_{32})+\xi(r_{21})\xi(r_{3})+\xi(r_2)\xi(r_{31})\\ &+ \Gamma(\pmb{r}_1,\pmb{r}_2)+\Gamma(\pmb{r}_1,\pmb{r}_3)+\Gamma(\pmb{r}_2,\pmb{r}_3)+\Gamma(\pmb{r}_{21},\pmb{r}_{31})\\ &+\Delta(\pmb{r}_1,\pmb{r}_2,\pmb{r}_3).
\end{split}
\end{equation}

Solving for the connected part of the four-point correlation function by retaining only the lowest order terms $O(\xi^3)$, we obtain
\begin{equation}
\begin{split}
\Delta(\pmb{r}_1,\pmb{r}_2,\pmb{r}_3) &= \xi(r_1)\xi(r_2)\xi(r_3) + \xi(r_1)\xi(r_2)\xi(r_{31})\\ &+ \xi(r_1)\xi(r_2)\xi(r_{32}) + 13 \textnormal{ perm.} + O(\xi^4).
\end{split}	
\end{equation}

The same equation is obtained in \cite{Joachimi11} in the context of the two-dimensional convergence field (eq. B11 in appendix B2). The Fourier transform of the four-point connected component $\Delta$ corresponds to the trispectrum $T(\pmb{k}_1,\pmb{k}_2,\pmb{k}_3)$. The trispectrum contribution to the covariance is obtained by considering its parallelogram configurations, yielding
\begin{equation}
\begin{split}
T(k_{1},k_{2}) &= P(k_1)P(k_2) \left[\phantom{\frac{|}{|}} 2 P(k_{1}) + 2 P(k_{2}) \right. \\&\left. + 2P(k_1+k_2)+2P(|k_1-k_2|) \right.\\ & \left. +\frac{P(k_1)}{P(k_2)}P(k_1+k_2) + \frac{P(k_1)}{P(k_2)}P(|k_1-k_2|) \right.\\ & \left. + \frac{P(k_2)}{P(k_1)}P(k_1+k_2) +\frac{P(k_2)}{P(k_1)}P(|k_1-k_2|) \right] + O(P^4).
\end{split}
\end{equation}

We obtained that the trispectrum contribution to the covariance of a lognormal field scales as $T(k_1,k_2) \propto P(k_1)P(k_2)f(k_1,k_2)$. If we speculate that $f(k_1,k_2)$ is approximately constant, this matches the non-Gaussian form of the $\alpha$-model covariance. While this far from a proof, it suggests that the ansatz of \cref{eq:covariance} may be justified in view of the exponential growth of structure. In fact, a more accurate fit to the covariance matrix measured from $N$-body simulations still can be written as $C(k_1,k_2) = P(k_1)P(k_2)g(k_1,k_2)$, namely a non trivial modulation of equation \cref{eq:covariance} \citep{Klypin18}.

\section{Application to the Displacement Field}\label{sec:lagscheme}
In this section we apply the ITAM simulation scheme to another random field, the divergence of the Lagrangian displacement field $\psi=\nabla \cdot \pmb{\Psi}$. This application not only serves as a proof of concept of the algorithm, but it is worth considering also in view of the relevance of $\psi$ to perform fast simulations of the matter density field \citep{Munari17}. Indeed, most of these schemes are based on approximations of $\psi$ to describe the evolution of dark matter particles in simulations. Thanks to ITAM, we can accurately reproduce $\psi$ measured from a simulation by retaining only its PDF and correlation structure, thus allowing us to elucidate the information content they bear compared with other semi-analytical techniques available in the literature. We first briefly summarize some of these analytical approximations for $\psi$.

\subsection{2LPT}
The evolution of preassureless particles evolving in an expanding Universe is given by the solution of \citep{Bernardeau01}
\begin{equation}\label{eq:fundamental}
	\frac{d^2}{d \tau^2} \pmb{x} + \mathcal{H} \frac{d}{d \tau} \pmb{x} = -\nabla \Phi(\pmb{x}),
\end{equation}
where we used $d \tau = dt / a$, $\mathcal{H}= H/a$. $H$ and $a$ represent the Hubble rate and the scale factor respectively, while $\Phi$ is the peculiar gravitational potential. In the Lagrangian picture of structure formation one solves for the displacement field $\pmb{\Psi}$, namely the mapping between Lagrangian coordinates $\pmb{q}$ onto Eulerian coordinates $\pmb{x}$:
\begin{equation}\label{eq:master}
	\pmb{x(q)} = \pmb{q} + \pmb{\Psi}(\pmb{q}).
\end{equation}

In Lagrangian perturbation theory one seeks a perturbative solution for \cref{eq:fundamental} of the form
\begin{equation}
	\pmb{\Psi}= \pmb{\Psi}^{(1)}+\pmb{\Psi}^{(2)}+ ...
\end{equation}

In practice we do not solve for the displacement directly, but we solve for its divergence $\psi^{(1)} = \nabla \cdot \pmb{\Psi}^{(1)}(\pmb{q},\tau)$ \citep{Buchert92, Bouchet1995}. Specifically, the famous ZA is the first-order solution %\cite{Zeldovich1970}
\begin{equation}\label{eq:zeld}
	\psi^{(1)} = -\delta^{(1)}(\pmb{q},\tau) = -\delta^{(1)}(\pmb{q}) D_1(\tau),
\end{equation}
where in the last equality we exploited the fact that the growth function $D_1$ at linear order is separable from the spatial part, and is given by the standard equation
\begin{equation}
D_1'' +\mathcal{H} D_1'= \frac{3}{2}  \mathcal{H}^2 \Omega_m(\tau) D_1(\tau).
\end{equation}

The ZA is often used to generate initial conditions of $N$-body simulations. Indeed, by having an initial Gaussian power spectrum we can generate a primordial density field $\delta^{(1)}(\pmb{q})$, which can be evolved to the actual displacement by integrating \cref{eq:zeld}. This is possible only under the assumption that the field is \textit{irrotational}, namely there exists a scalar potential $\phi^{(1)}(\pmb{q})$ such that
\begin{equation}
\pmb{\Psi}^{(1)} = -\nabla \phi^{(1)}.
\end{equation}

The assumption of irrotationality has been studied thoroughly by \cite{Chan2014}, and has been confirmed to be accurate up to $k \sim 1 \ h/Mpc$, even at $z=0$. To increase the precision of the initial conditions, the perturbative solution is found at higher orders, for example at second order:
\begin{equation}\label{eq:2lpt}
	\psi^{(2)}(\pmb{q},\tau) = \frac{D_2(\tau)}{2 D^2_1(\tau)} \sum_{i \neq j} \left( \phi^{(1)}_{i,i} \phi^{(1)}_{j,j} - \phi^{(1)}_{i,j} \phi^{(1)}_{j,i}\right),
\end{equation}
where $D_2$ is the second order growth factor. Fitting functions for the growth factors are respectively \citep{Carrol1992}
\begin{equation}
	D_{1}(a) \simeq \frac{5}{2} \frac{a \Omega_m(a)}{\Omega_m(a)^{4/7}-\Omega_{\Lambda}(a)+(1+\Omega_m(a)/2)(1+\Omega_{\Lambda}(a)/70)},
\end{equation}
and \citep{Bouchet1995}
\begin{equation}
	D_2 \simeq -\frac{3}{7} D^2_1.
\end{equation}

Also for the second order we can assume irrotationality, which allows us to define the second order displacement potential 
\begin{equation}
    \pmb{\Psi}^{(2)} \equiv \nabla \phi^{(2)}.
\end{equation}

To summarize, the displacement in 2LPT is:
\begin{equation}
	\pmb{x} = \pmb{q} - \nabla \phi^{(1)}(\pmb{q},\tau) + \nabla \phi^{(2)}(\pmb{q},\tau).
\end{equation}

2LPT is ubiquitous in implementations of fast simulation schemes, being the basis for constrained realizations of density field \citep{Kitaura13,Jasche13,Leclercq15}, for fast mock catalogue realizations \citep{Monaco13,Stein18}, or hybrid approaches \citep{Tassev13}.

\subsection{\muscle}
Despite its success in the perturbative regime, 2LPT becomes inaccurate when density perturbations grow large, at high resolution and low redshift \citep{Matsubara08a}.
%Also, when shell crossing occurs, trajectories overlap, and the very same equations of motion cease to be valid.  In this limit particles tend to accumulate in excess, forming pancakes \citep{Bouchet1995}. -- True, but shell crossing is not properly treated in anything but N-body
In this non-perturbative regime, 2LPT produces extreme particle overcrossings at high density, and also overestimates densities at low densities, even producing overdense clumps at what should be extreme underdensities  \citep{Sahni1996, Neyrinck13}.

An improvement over the ZA and 2LPT approaches was proposed by \cite{Neyrinck13}. He adapted the Eulerian spherical approximation formula introduced by \cite{Bernardeau1994b}, to a form previously found by \citep{Mohayaee1996}, for the evolution of isolated spherical perturbations to describe the Lagrangian evolution of particles:
\begin{equation}\label{eq:sc}
    \psi_{sc}(\tau) =  3 \left[ \left(1-\frac{\delta^{(1)}(\tau)}{\gamma} \right)^{\gamma/3} -1 \right].
\end{equation}

The parameter $\gamma$ here is related to the spherical-collapse density; following e.g.\ \citet{Neyrinck13}, we adopt a value $\gamma=3/2$. This formula belongs to the broader family of local Lagrangian mappings \citep{Protogeros97}, as it links the value of the density at position $\pmb{q}$ and time $\tau$ with the new value at the same position in Lagrangian coordinates. The spherical collapse formula improves the cross correlation with $N$-body simulation at the level of smaller scales, but it fails to reproduce the large scale displacements of 2LPT.   This impediment was overcome by \cite{Kitaura13_ALPT}, with Augmented Lagrangian Perturbation Theory (ALPT). Here, an interpolating smoothing kernel is a applied to the divergence displacement in order to combine the spherical collapse at small scales with 2LPT at large scales. This method requires a fitting parameter for this separation scale. Providing an alternative, \cite{Neyrinck15} proposed a non-perturbative, parameter-free technique that has a similar performance to ALPT. This approach is called MUltiscale Spherical ColLapse Evolution (\muscle), where the condition for collapse is checked in the initial density field on increasingly larger smoothing scales. If at some smoothing radius $R_c$ it occurs $\delta(\pmb{q},R_c)>\gamma$, $\psi(\pmb{q})=-3$ is set at that voxel, indicating a halo particle as it is observed in $N$-body simulations \citep{Neyrinck13}. The multi-scale approach of \muscle\ accounts for the void-in-cloud problem, which leads to retrieving the large scale displacement, and can be summarized as follows
\begin{equation}\label{eq:muscle}
\psi_{msc}(\pmb{x})=
\begin{cases}
3\left[ \left(1-\frac{\delta_l}{\gamma} \right)^{\gamma/3} -1. \right] \ \ \ \ \delta_l(R)<\gamma, \ \forall R, \\
\\
-3, \ \ \ \ \ \ \ \ \ \ \ \ \ \ \ \ \ \ \ \ \ \ \ \ \ \ \ \ \ \ \ \ \  \delta_l(R_{c}) > \gamma.
\end{cases}
\end{equation}

We have updated the previous code for these schemes, putting it in a new github repository\footnote{\url{https://github.com/tos-1/MUSCLE}}.
\begin{figure}
	\includegraphics[width=0.45\textwidth]{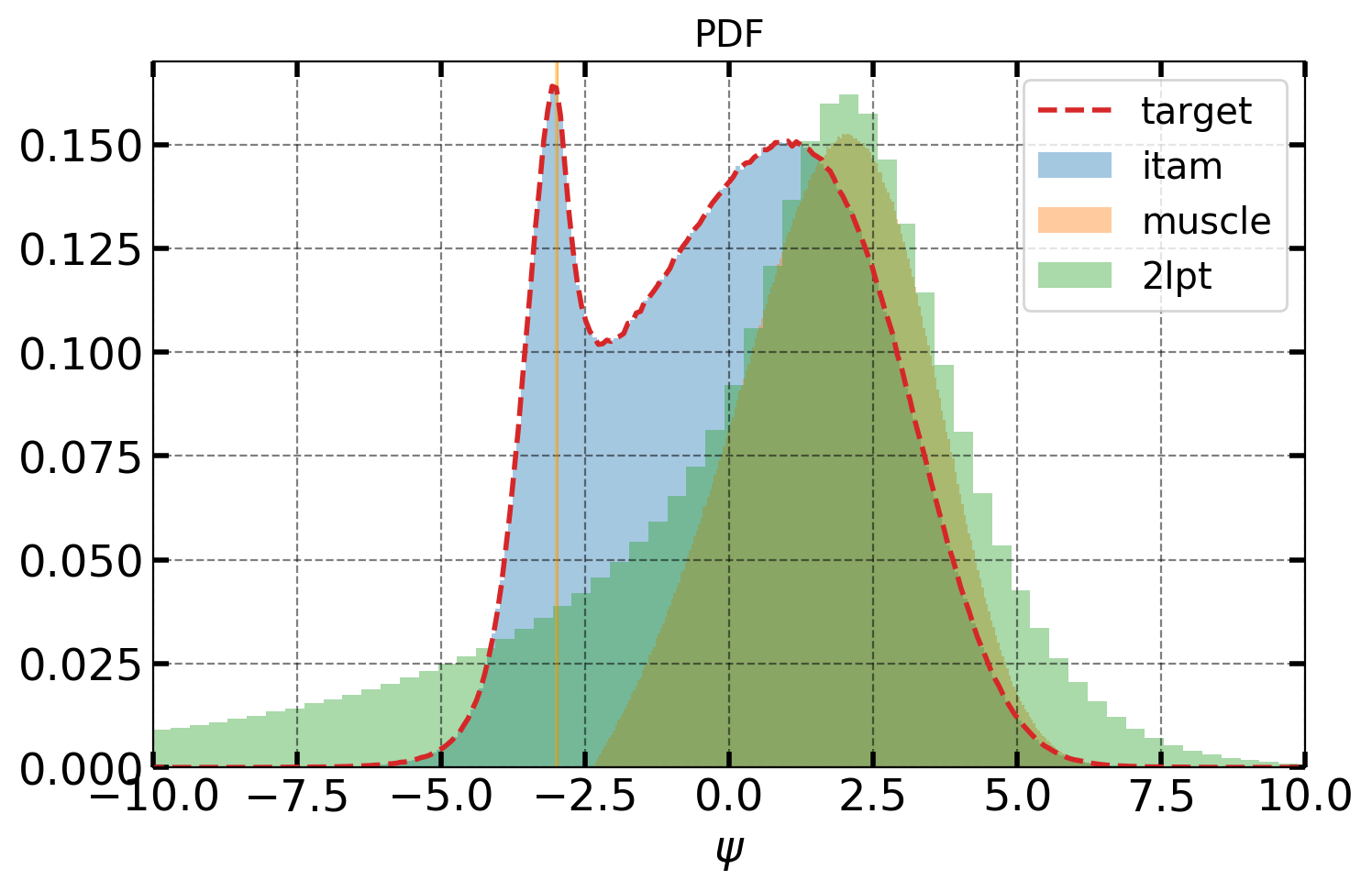}\caption{PDFs of $\psi$ for various schemes. The peak at $\psi \sim -3$ is associated with collapsed halo particles. In \muscle, halo particles give a Dirac-delta function at $\psi = -3$ (see \cref{eq:muscle}). \label{fig:psi_PDFs}}
\end{figure}
\begin{figure}
	\includegraphics[width=0.45\textwidth]{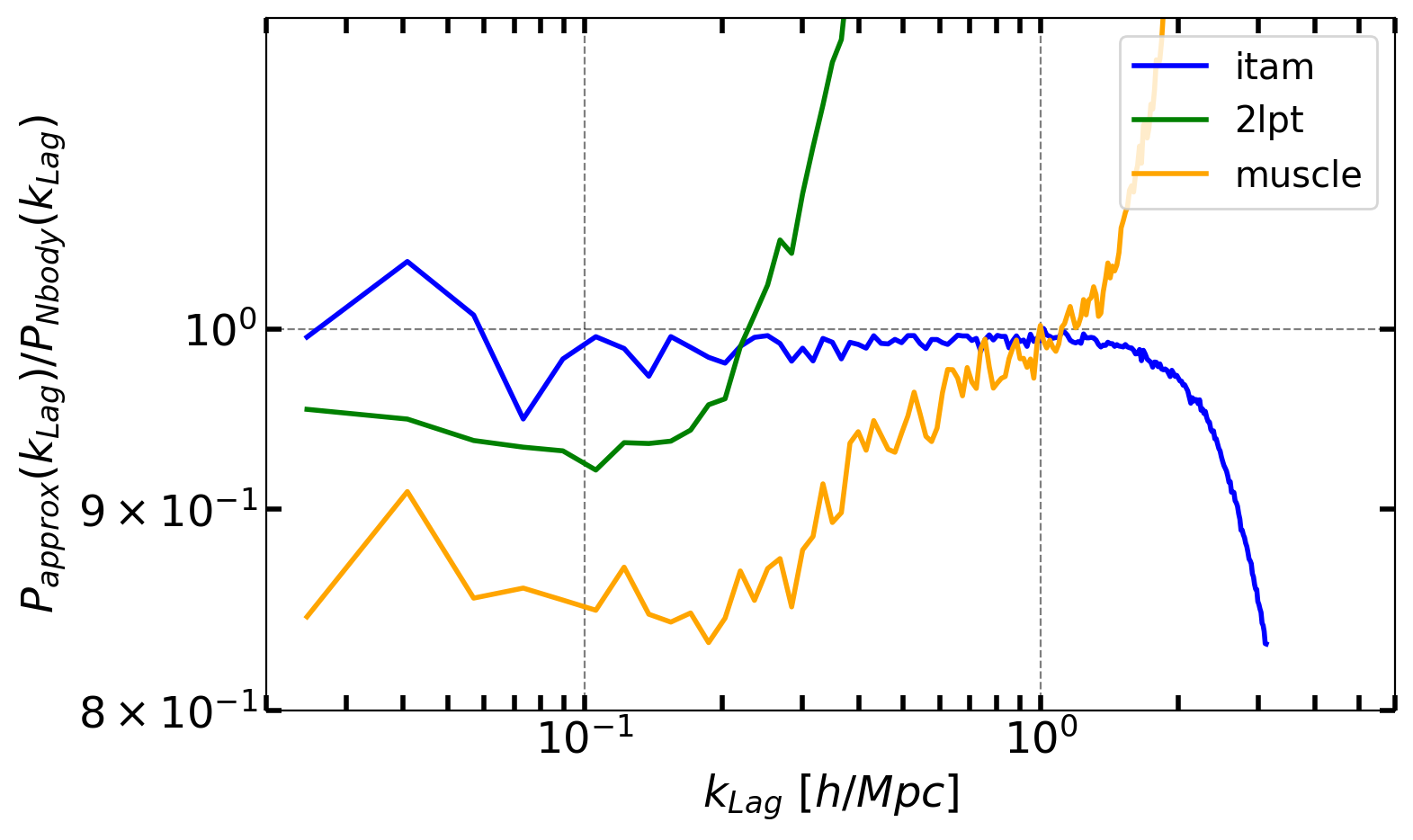}\caption{ Transfer function of $\psi$, defined as the ratio of the power spectrum of $\psi$ generated from the fast simulation scheme over the one measured from the $N$-body realization. \label{fig:psi_transfer}}
\end{figure}
\begin{figure}
	\includegraphics[width=0.45\textwidth]{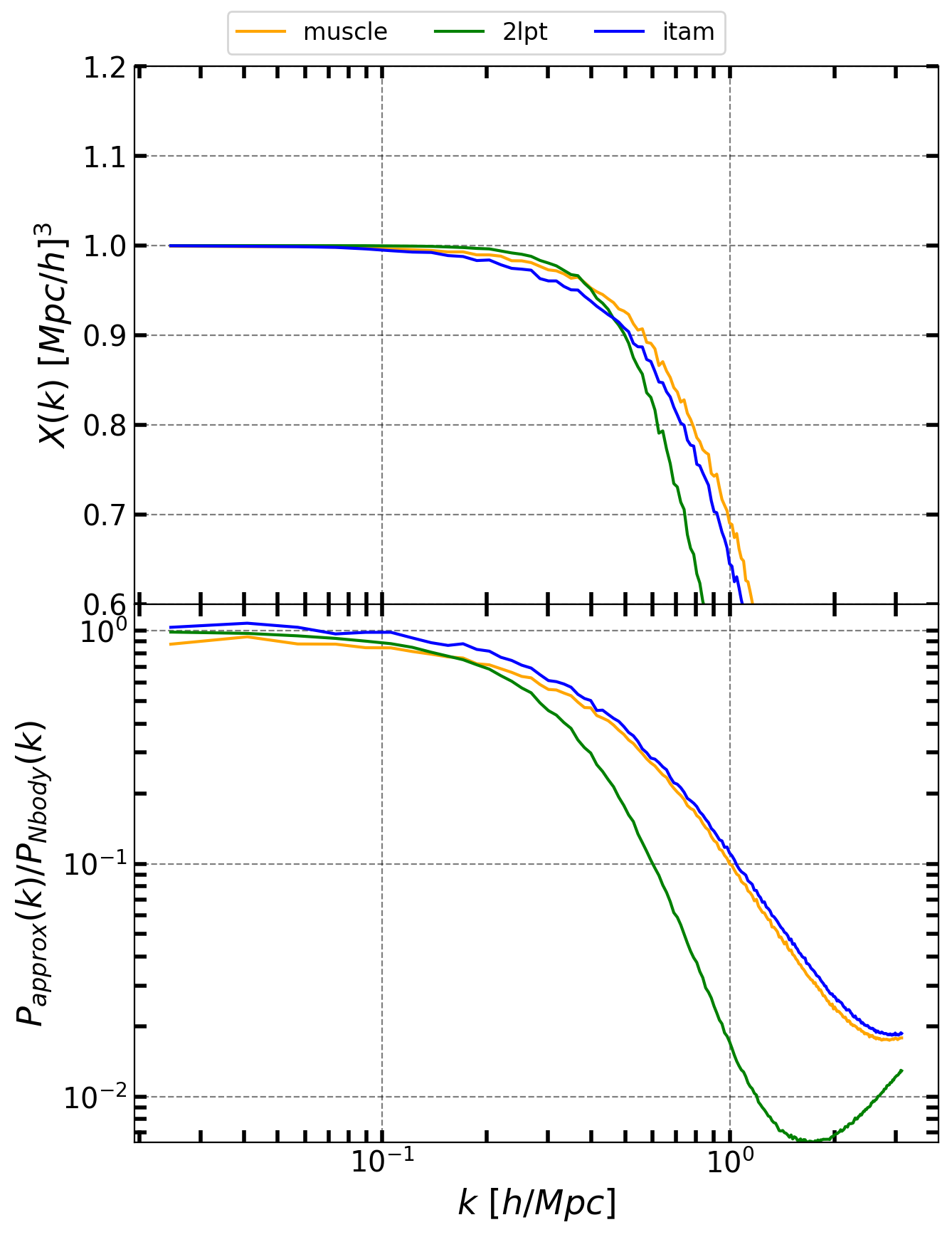}\caption{ The top panel shows the cross correlation between the full-gravity matter field and those resulting from the approximations. The bottom panel shows the transfer function of the matter power spectra. \label{fig:psi_crossntransfer}}
\end{figure}
\begin{figure}
		\includegraphics[width=0.45\textwidth]{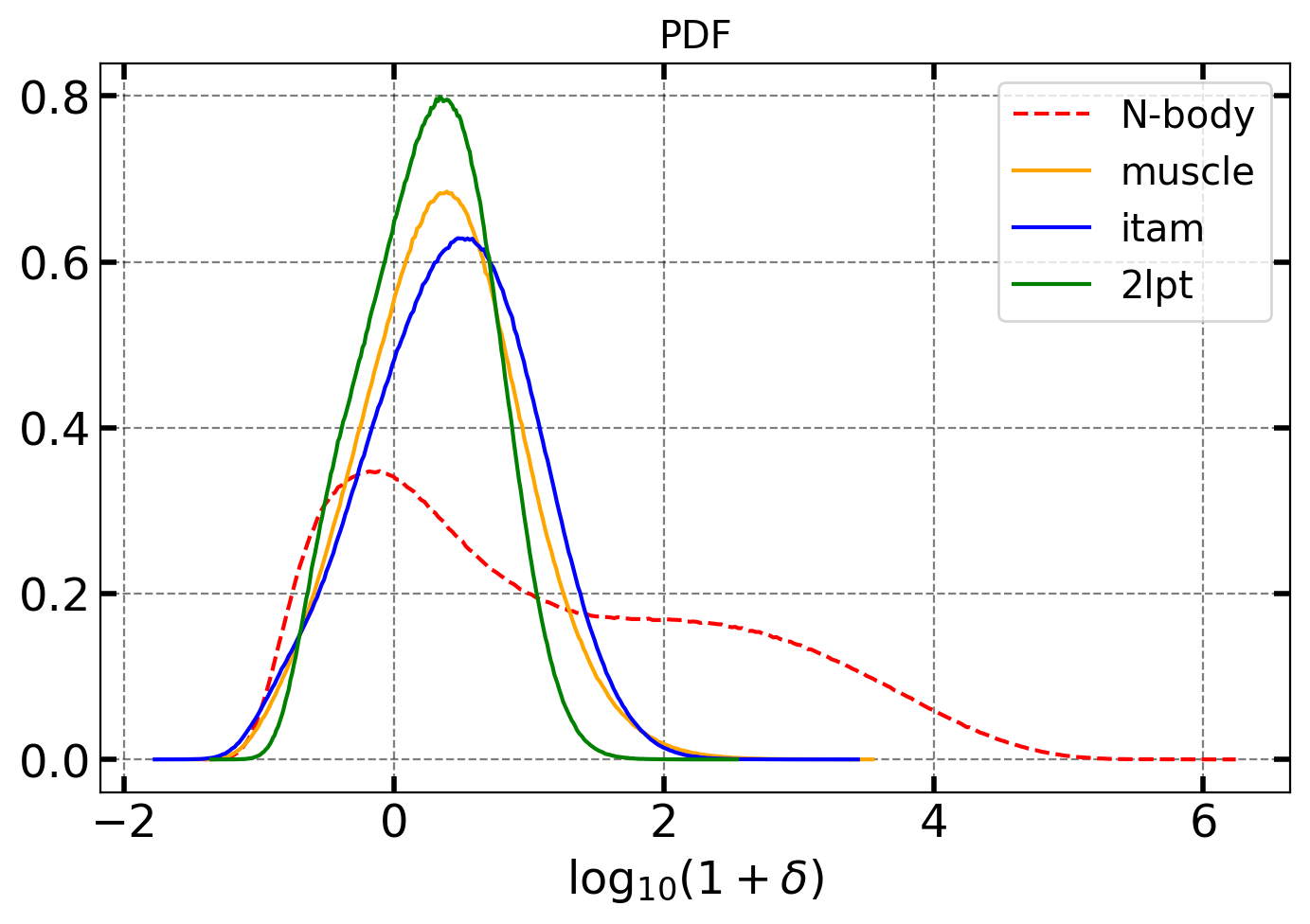}\caption{ PDF of the matter fields generated by the divergence displacement fields listed in the legend. These densities have been measured with the Voronoi tessellation code in \textsc{voboz/zobov}. \citep{Neyrinck05}\label{fig:voboz_psi_densities_PDFs}}
\end{figure}

\subsection{Translation Process for Lagrangian Simulations}
Ultimately, all previous techniques map the initial density field onto the displacement divergence field without explicitly depending on the coordinates. This kind of mapping is a local Lagrangian mapping \citep{Protogeros97}, and it falls in the same category of the mappings considered by translation process theory, being a local nonlinear transformation. More precisely, the ZA and spherical collapse mappings (\cref{eq:zeld,eq:sc}) are exactly monotonic, while 2LPT and \muscle\ (\cref{eq:2lpt,eq:muscle}) use non-local information. In light of translation process theory, we could attempt to optimize for the pre-translation Gaussian field that is most adapted to be transformed onto $\psi$, and we can check whether this yields an improvement in encoding the evolution of the matter density field with respect to these techniques.   In order to generate an ITAM realization of $\psi$, we measure the target power spectrum $\langle \psi(\pmb{k}) \psi^{*}(\pmb{k}') \rangle \equiv (2 \pi)^3 P_{\psi}(k) \delta(\pmb{k}+\pmb{k}')$, and $\psi$'s PDF, from the SB simulation. In using both from a single realization, the results may depend on peculiarities in that realization. The limitation is due to the lack of an available software to produce accurate fitting functions for the power spectrum and PDF of $\psi$. This introduced dependence is not a problem, as long as one has the aim of isolating the contribution from the PDF and the correlation of $\psi$ to the specific realization examined, to assess how much information they convey with respect to the exact result. To emulate the final $\psi$ field, we set the phases of the pre-translation Gaussian field to the phases of the initial density field of the realization.
%For consistency, the pre-translation Gaussian field produced by ITAM must share the same phases as the initial density field of the $N$-body, before transforming it onto $\psi$. -- I'm not sure why this is "for consistency", and why we "must" do it ... but it is definitely reasonable
We do not smooth the measured $\psi$ field, because both the power spectrum and the PDF are measured at the grid level, and one can check that the $\psi$ field variance is already consistent with its power spectrum variance integrated up to the Nyquist frequency.
%We do not smooth the measured $\psi$ field, because, contrary to the matter density examined before, this is a Lagrangian quantity that is measured directly at the grid level.

In \cref{fig:psi_PDFs}, we can see that the PDF of the $\psi$ realization from ITAM matches the one of the $N$-body simulation (much better than \muscle, and particularly 2LPT), as it should by the virtue of the translation transform \cref{eq:g}. Likewise, the Lagrangian-space $\psi$ power spectrum of ITAM matches that of the $N$-body simulation better than the other prescriptions, as shown in \cref{fig:psi_transfer}. The two standard simulation schemes have a mismatch, with 2LPT diverging the most at small scales, and \muscle\ departing the most on large scales. This confirms the detailed analysis of the power spectra of Lagrangian schemes conducted by \cite{Chan2014}, where it is also found that these techniques come short of matching the power spectrum measured in $N$-body simulations. Another benchmark of the goodness of a fast simulation scheme is the similarity between the resulting matter density and the $N$-body matter density. This is generally quantified by means of the cross correlation spectrum, that for two fields A and B is 
\begin{equation}\label{eq:cross}
	X(k) = \frac{ \langle \delta_{A}(k)\delta^*_{B}(k) \rangle }{P_A{(k)} P_B{(k)}}.
\end{equation}

\begin{figure*}
	\includegraphics[width=0.85\textwidth]{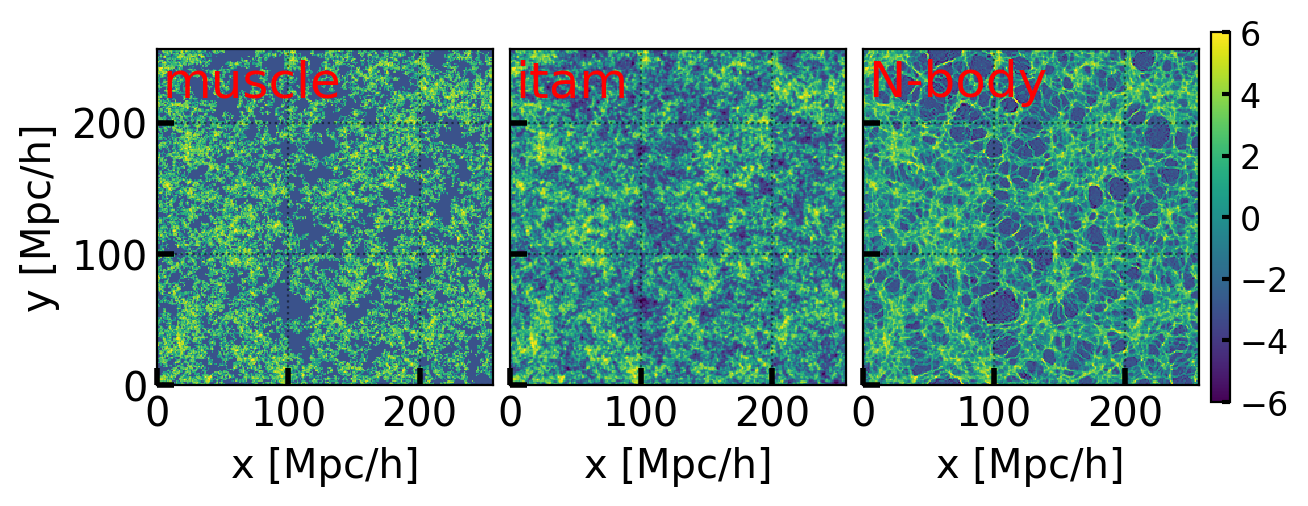}
	
	\caption{A 1-\hmpc -thick Lagrangian slice showing the $\psi$ field in various schemes. \muscle\ performs better than ITAM in locating the patches of haloes in Lagrangian space. \label{fig:psi_densities}}
\end{figure*}

\begin{figure*}
	\includegraphics[width=0.7\textwidth]{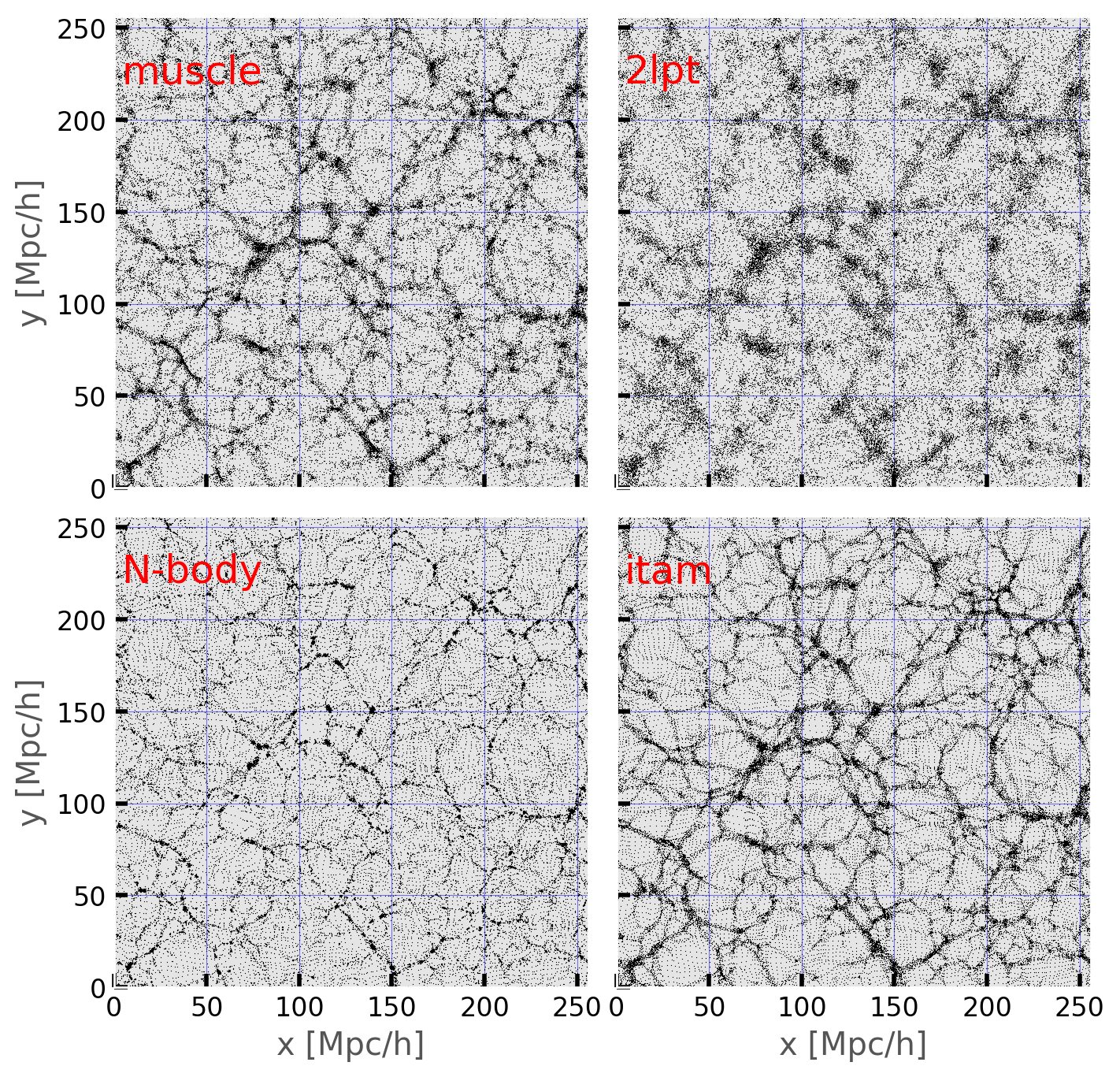}
	\caption{A 1-\hmpc -thick Lagrangian slice showing the Eulerian position of particles for the simulation schemes considered in the text. ITAM performs better than \muscle\ in collapsing filaments and walls. \label{fig:lagslice}}
\end{figure*}

The cross correlations are plotted in the top panel of \cref{fig:psi_crossntransfer}. As we can see, the \muscle\ implementation still performs better than ITAM, which in turn is better than 2LPT, at least on smaller scales. Examining this further, 2LPT is almost a local approach, in the sense that it considers the information of each voxel and its nearest neighbours (\cref{eq:2lpt}). ITAM by construction is non-local, as it involves the two-point correlation function. \muscle\ is also non-local, but the non-locality is coming from a smoothing procedure on various scales centered around each voxel (\cref{eq:muscle}). Considering that the one-point mapping of ITAM is very well approximated by the spherical collapse formula (e.g. fig. 5 of \cite{Neyrinck15}), it seems that the multi-scale information is more relevant than the non-local information of the two-point correlation function. In fact from \cref{fig:psi_densities} we can see that halo particles are more accurately tagged by \muscle, rather than ITAM. However, when one examines the particle positions plotted in \cref{fig:lagslice}, it is apparent that ITAM is collapsing more filaments and walls when compared to \muscle.
%which reflects the better description of the correlation function. -- I'm not sure this follows
This improvement appears also in the bottom panel of \cref{fig:psi_crossntransfer}. The matter power spectrum of density field generated through ITAM on $\psi$ is closer to the $N$-body case. In \cref{fig:voboz_psi_densities_PDFs}, ITAM also manages to capture the highest densities of any approximation, quantifying the high filament compactness that ITAM manages in \cref{fig:lagslice}. Overall, it seems that the improvement, despite not being substantial, could pave the way for a new and more accurate fast simulation scheme, as e.g. a combination of \muscle\ and ITAM. We leave this investigation for a future study.

\section{Summary \& Conclusions}\label{sec:Conclusions}
In this work we have adapted the ITAM algorithm, first suggested by \cite{Shields11} for one-dimensional time simulations, to the three-dimensional case of spatial fields (\cref{fig:FlowChart}). ITAM optimizes for an effective Gaussian power spectrum to generate many realizations of a field that can be mapped onto the target field under a simple monotonic transform. The target field produced in this way has prescribed PDF and power spectrum, which are the only required inputs of the code. Using as a reference the PDF and the power spectrum of an $N$-body simulation, we are able to generate many realizations of a nonlinear field that match the density field at least at the level of the one-point and two-point correlation functions, thus improving over the usual lognormal assumption. We compared the covariance of the ITAM field against the lognormal field, showing that the accuracy of ITAM is comparable to the lognormal case, or even closer to the expected covariance (\cref{fig:diagonal,fig:off_diagonal}). For the first time, we also motivate analytically a common ansatz used for a phenomenological description of the covariance matrix, showing that its form is a result of the Eulerian growth of structure such as an exponential growth in the case of a lognormal field (\cref{subsec:bias}). In fact, this functional form is found directly from examining the trispectrum of the lognormal field.    As a last example, we apply ITAM to the case of the bi-modal PDF of the Lagrangian displacement field (\cref{fig:psi_PDFs}). As expected, we find that the displacement field generated by ITAM is more faithful to the exact one measured from the $N$-body simulation than available semi-analytical schemes (\cref{fig:psi_transfer}), as well as being effective in generating a more accurate density field (\cref{fig:voboz_psi_densities_PDFs,fig:lagslice}). This result suggests that further improvements for Lagrangian schemes are not expected to come from one-point mappings of the initial density field, but are to be found in multi-scale or non-local approaches.

\section*{Acknowledgements}
FT acknowledges financial support by ASI Grant No. 2016-24-H.0 and thanks the "Dipartimento di Fisica Aldo Pontremoli" of the University of Milano, and the Department of Theoretical Physics at the University of the Basque Country in Bilbao for hospitality during the development of this work. 
LG, BRG and FT acknowledge financial support by grant MIUR PRIN 2015 "Cosmology and Fundamental Physics: illuminating the Dark 
Universe with Euclid". MCN is grateful for funding from Basque Government grant IT956-16. FT is grateful to Michael D. Shields for clarifications on ITAM and acknowledges useful discussions with Raul Angulo, Cora Uhlemann, Enzo Branchini and Carmelita Carbone.

\section*{Data availability}
The codes for ITAM and \muscle\ simulation schemes are available at \url{https://github.com/tos-1/ITAM} and \url{https://github.com/tos-1/MUSCLE} respectively. 

\bibliographystyle{mnras}
\bibliography{StochasticProcesses}
%
%\appendix*

% Don't change these lines
\bsp	% typesetting comment
\label{lastpage}
\end{document}